\documentclass[apj]{emulateapj}
\usepackage{apjfonts}
\usepackage{amsbsy}

\def\slogm{\sigma_{{\rm log}M}}
\def\om{\Omega_m}

\def\omb{\Omega_b}
\def\s8{\sigma_8}

\def\etal{et\,\,al.}
\def\hmpc{$h^{-1}\,$Mpc}

\def\xg{$\xi_g(r)$}

\def\x2{$\chi^2$}
\def\hmsol{$h^{-1}\,$M$_\odot$}

\def\wp{$w_p(r_p)$}

\def\cmbfast{{\scriptsize CMBFAST}}

\def\kmsmpc{km\,s$^{-1}$\,Mpc$^{-1}$}

\def\plin{P_{\rm lin}(k)}

\def\mstar{M_\ast}
\def\mmin{M_{\rm min}}

\def\mcut{M_{\rm cut}}
\def\sigmaM{\sigma_{\log M}}
\def\navg{\langle N\rangle_M}

\def\nsat{\langle N_{\mbox{\scriptsize sat}}\rangle_M}

\def\ncen{\langle N_{\mbox{\scriptsize cen}}\rangle_M}

\def\ngavg{\bar{n}_g}

\def\NNm1{\langle N(N-1) \rangle}

\def\xid{\xi_{\rm 2h}}

\def\dc{\delta_c}
\def\ds{\delta_s}
\def\fmin{f_{\rm min}}
\def\fsat{f_{\rm sat}}

\begin{document}

\title{Cosmic Voids and Galaxy Bias in the Halo Occupation Framework}

\author{Jeremy L. Tinker\altaffilmark{1}, 
David H. Weinberg\altaffilmark{2}, \&
Michael S. Warren\altaffilmark{3}}

\altaffiltext{1}{Kavli Institute for Cosmological Physics, Department
  of Astronomy \& Astrophysics, The University of Chicago, IL 60637}

\altaffiltext{2}{Department of Astronomy, The Ohio State University,
140 W. 18th Avenue, Columbus, Ohio 43210} 

\altaffiltext{3}{Theoretical Astrophysics, Los Alamos National Laboratories, 
P.O. Box 1663, Los Alamos, New Mexico, 87545}

\begin{abstract}

We investigate the power of void statistics to constrain galaxy bias and the
amplitude of dark matter fluctuations.  We adopt a $\Lambda$CDM
cosmological scenario (inflationary cold dark matter with a cosmological
constant) and use the halo occupation distribution (HOD) framework to describe
the relation between galaxies and dark matter. After choosing HOD parameters
that reproduce the mean space density $\ngavg$ and projected correlation
function \wp\ measuremed for galaxy samples with $M_r<-19$
and $M_r<-21$ from the Sloan Digital Sky Survey (SDSS), we predict the void 
probability function (VPF) and underdensity probability function (UPF) of these
samples by populating the halos of a large, high-resolution N-body simulation
(400\hmpc, $1280^3$ particles). If we make the conventional simplifying
assumption that the HOD is independent of large scale environment at fixed halo
mass, then models constrained to match $\ngavg$ and \wp\ predict nearly
identical void statistics, independent of the scatter between halo mass and
central galaxy luminosity or statistical uncertainties in HOD parameters.
Models with power spectrum normalizations
$\sigma_8=0.7$ and $\sigma_8=0.9$ also predict very similar void
statistics, with stronger galaxy bias compensating for weaker dark matter
fluctuations in the low-$\sigma_8$ model.
However, the VPF and UPF are sensitive to environmental variations
of the HOD in a regime where these variations have little impact on \wp.
For example, doubling the minimum host halo mass in regions with large scale
(5\hmpc) density contrast $\delta<-0.65$ has a readily detectable impact
on void probabilities of $M_r<-19$ galaxies, and a similar change for 
$\delta<-0.2$ alters the void probabilities of $M_r<-21$ galaxies at
a detectable level.  The VPF
and UPF provide complementary information about the onset and magnitude of
density-dependence in the HOD. By detecting or ruling out HOD changes in low
density regions, void statistics can reduce systematic uncertainties in the
cosmological constraints derived from HOD modeling, and, more importantly, 
reveal connections between halo formation history and galaxy properties.

\end{abstract}


\section{Introduction}

Voids are an omnipresent feature of the galaxy distribution
(\citealt{kirshner87, delapparent91, vogeley94, hoyle04, conroy05};
see \citealt{rood88} for a historical review). Galaxy redshift surveys
reveal that underdense regions 30 \hmpc\ in diameter are common, and
that they can be as large as 50 \hmpc. A fundamental question
regarding the nature of voids is whether they are as empty of matter
as they are of light, or whether the breadth and depth of observed
voids is a consequence of a lower efficiency of galaxy formation in
underdense regions. For the matter distribution, the frequency of
large voids is sensitive to departures from Gaussianity in the
primordial density fluctuations (\citealt{weinberg92}). In Gaussian
models, the frequency of voids depends on the rms amplitude of matter
clustering and on the bias between galaxies and mass
(\citealt{einasto91, weinberg92, little94}). \cite{little94} show that
void statistics are sensitive not only to the amplitude of galaxy
bias, but also to the detailed form of the relation between the
underlying matter distribution and galaxy formation efficiency. In
this paper, we revisit the connection between voids and galaxy bias in
the framework of the halo occupation distribution (HOD), which has
emerged in the last few years as a powerful method for characterizing
the relationship between galaxies and mass (\citealt{kauffmann97,
  benson00, seljak00, peacock00, ma00, roman01, bw02}, hereafter BW).

Many different methods have been proposed for identifying voids and
quantifying their importance (\citealt{ryden84, kauffmann91,
  kauffmann92, el-ad97, hoyle02}). In this paper we will use two of
the simplest statistical measures of cosmic voids: the void
probability function (VPF, denoted $P_0$) and the underdensity
probability function (UPF, denoted $P_U$).  The VPF is defined as the
probability that a randomly placed sphere of radius $r$ contains no
galaxies of a given class. The UPF is similar but quantifies the
probability that the galaxy density within a randomly placed sphere is
less than a defined threshold, usually 0.2 times the mean density for
galaxies of that type. The VPF depends on the full hierarchy of
$n$-point correlation functions (\citealt{white79}), so in principle
it contains complementary information to the two-point correlation
function \xg. The first significant measurement of the VPF and UPF was
the analysis of the CfA redshift survey by \cite{vogeley92,
  vogeley94}.  The current state of the art is the analysis of the
Two-Degree Field Galaxy Redshift Survey (2dFGRS, \citealt{colless01})
by \cite{hoyle04}, which covered 1500 deg$^2$ to a median redshift of
$\sim 0.1$. Forthcoming results for the Sloan Digital Sky Survey
(SDSS, \citealt{york00}) should yield even more precise measurements
of the void distribution owing to its larger survey volume, which will
ultimately extend over nearly a quarter of the sky.

The galaxies that define the voids are, in general, a biased tracer of
the underlying mass distribution. A fully non-linear model of galaxy
bias is required to compare theory to observations for any clustering
measure. The HOD formalism describes bias at the level of dark matter
``halos,'' where a halo is a structure of overdensity
$\rho/\bar{\rho}\sim 200$, in approximate dynamical equilibrium
(identified in this paper by the Friends-of-Friends algorithm, see
\S\ref{sec:sims}).  With this definition, the mass function, spatial
clustering, and velocity statistics of dark matter halos are
essentially independent of gas physics, and they can be determined
accurately for a given cosmology using high-resolution N-body
simulations (e.g., \citealt{warren05, springel05}) or analytic
approximations calibrated on these simulations.  In the HOD framework,
one specifies the bias of any particular class of galaxies by
specifying $P(N|M)$, the probability that a halo of mass $M$ houses
$N$ such galaxies, together with any spatial or velocity biases
between galaxies and dark matter {\it within} individual halos.  To
the extent that the HOD is independent of a halo's larger scale
environment (a point that we return to below), this description of
bias is complete, for all clustering statistics on all scales.  The
HOD framework has been used to model measurements of the projected
galaxy correlation function, \wp, from the SDSS (\citealt{zehavi04a,
  zehavi04b, tinker05a}). The conditional luminosity function, which
combines the HOD with luminosity information, has been used to analyze
the correlation function and luminosity function of the Two-Degree
Field Galaxy Redshift Survey (e.g. \citealt{yang03, yang04,
  vdb03}). Approaches to modeling void statistics with the HOD have
been discussed by \cite{benson01} and BW. Here we follow the approach
of BW and calculate the VPF and UPF by populating dark matter halos
identified in collisionless N-body simulations.

It is well known that the halo mass function correlates with
large-scale environment (\citealt{bond91, lemson99, sheth02}). Massive
halos form from high-sigma fluctuations in the primordial density
field, which reside preferentially in regions that are overdense on
larger scales \citep{kaiser84}.  In low-density regions, high-sigma
fluctuations are rare, so these environments are dominated by
low-mass halos. We therefore expect void statistics to be sensitive to
halo occupation at low halo masses, where the occupation number is
zero or one. BW adopted a power-law mean occupation function for
galaxies above a luminosity threshold, i.e., $\navg \equiv \sum
NP(N|M) = (M/M_1)^\alpha$, with a cutoff at $\mmin$ below which halos
could not house galaxies above the threshold. Their results
showed a strong correlation between $\mmin$ and the sizes of
voids. Here we adopt a more physically motivated occupation function
that divides halo occupation into central and satellite galaxies
(\citealt{guzik_seljak02, kravtsov04, zheng04b}). In this
parameterization, the behavior of $\navg$ at low masses is essentially
determined by the relation between halo mass and the properties of the
central galaxy. For each galaxy population that we model, we only
consider HOD models that are constrained to match the mean space
density $\ngavg$ and the projected correlation function \wp\ of SDSS
galaxies measured by \cite{zehavi04b}. In this respect, we are
investigating what {\it additional} information can be obtained from
void statistics once these constraints have been imposed. In this
context, we want to answer several questions:

1. Do void statistics provide additional information on the minimum
halo mass for a galaxy sample, within the range allowed by the
constraints of matching the observed $\ngavg$ and \wp?

2. Are void statistics sensitive to the scatter in the relation between
halo mass and central galaxy luminosity, which alters the {\it form} of
the cutoff in $\navg$?

3. Can void statistics discriminate among values of $\s8$, the rms
linear theory amplitude of {\it mass} fluctuations in 8
\hmpc\ spheres, if {\it galaxy} HODs are chosen to reproduce the
observed correlation function? (Here $h\equiv H_0/100$ \kmsmpc .)

We will show that the answer to all three questions is, for practical
purposes, no. Once one has matched $\ngavg$ and \wp, variations in the
VPF and UPF that result from varying $\mmin$ within the allowed range,
from changing the form of the low-mass cutoff, or from altering $\s8$,
are all at the level of error expected from the full SDSS sample. We
will demonstrate that the aspect of the HOD that has the greatest
influence on void statistics is the fraction of galaxies that are
satellites, because once this is determined the cutoff mass $\mmin$ is
essentially fixed by $\ngavg$. The satellite fraction is well
constrained by matching \wp. Therefore, within the standard
cosmological and HOD framework, the predicted void statistics are
remarkably robust. The imposition of the observed
correlation function as a constraint is the crucial difference
between our work and the analysis of BW, and it accounts for our
differing conclusions about the information content of void statistics.

In the standard implementation of HOD models, including those that
lead to the conclusions mentioned above, the statistics of galaxy
occupation are assumed to depend on a halo's mass, independent of its
larger scale environment. Semi-analytical models of galaxy formation
that incorporate this assumption reproduce the observed correlations
of galaxy luminosity, color, and morphology with environment
(\citealt{benson03, norberg01}); these arise because the halo mass
function itself varies with large scale overdensity. Direct
observational support for this assumption comes from the work of
\cite{blanton04}, who show that the variation of blue galaxy fraction
with overdensity measured on a 6 \hmpc\ scale is fully explained by
the variation on the 1 \hmpc\ scale characteristic of individual large
halos. On the theoretical side, \citeauthor{bond91}'s
(\citeyear{bond91}) excursion set derivation of the extended
Press-Schechter (1974) formalism predicts that halo formation
histories at fixed mass are independent of large scale overdensity
(\cite{white96}), while \cite{berlind03} show that the HOD for a
baryon-mass selected sample from Weinberg et al.'s (2004) smoothed
particle hydrodynamic simulation has no detectable dependence on
environment. However, while early N-body investigations
(\citealt{lemson99, sheth04}) showed little or no correlation between
halo formation histories and environment for masses above $M\sim
10^{13}$ \hmsol, several recent studies (\citealt{gao05, harker05,
  wechsler05, zhu06}) show much stronger correlations at lower
masses. The correlation of halo formation time with large scale
overdensity, which is especially prominent in the mass range $M\sim
10^{11}$ \hmsol\ of individual dwarf galaxy halos, raises the
possibility that the galaxy HOD itself varies with halo environment,
at least in the single-galaxy regime.

Void statistics should be a sensitive probe of any such environmental
variations, since they characterize extreme environments and are not
washed out by the strong signals from multiple galaxy halos. In \S 4, we
investigate a simple class of extended HOD models in which the minimum
host mass $\mmin$ changes by a fixed factor in low density
environments. The VPF and UPF prove quite sensitive to such changes,
and the two statistics provide somewhat complementary information
about the density dependence of $\mmin$. By tightly constraining this
dependence, void statistics can test fundamental ideas about galaxy
formation and provide essential input to cosmological applications of
HOD methods (\citealt{zheng05}).


\section{Numerical Simulations and HOD Models}
\label{sec:sims}

The simulation used in this paper was performed using the Hashed
Oct-Tree code of \cite{warren93}, and is similar to those presented in
\cite{seljakwarren04} and \cite{warren05}. This simulation is
substantially larger in terms of particle number, with $1280^3$
particles. The simulation box is 400 \hmpc\ per side, giving a mass
resolution of $2.5\times 10^9$ \hmsol.  Halos are identified in the
simulation by the friends-of-friends technique (\citealt{davis85})
with a linking length 0.2 times the mean interparticle separation. For
low-luminosity, $M_r\sim -19$ galaxies, the minimum mass halo roughly
corresponds to a 100-particle halo. The linear power spectrum used to
create the initial conditions was calculated with
\cmbfast\ (\citealt{cmbfast}), with the parameter set
$(\om,\Lambda,\s8,h,\omb) = (0.3,0.7,0.9,0.7,0.04)$ and a
scale-invariant inflationary power spectrum. We use the output at
$z=0.459$, where the linear growthfactor is $0.79$, to represent a
universe with $\s8=0.7$, scaling the peculiar velocities and internal
dispersions of halos to values appropriate for $\om=0.3$ (see Tinker
et.\ al. [2005b] and Zheng et.\ al [2002] for a discussion of this
technique).

This simulation is ideal for our purposes because of its high mass
resolution and large box size. High mass resolution is required to
model low-luminosity galaxy populations. The lowest luminosity SDSS
sample that we can model is $M_r<-19$.\footnote{All absolute magnitudes
  are quoted for $h=1$ throughout the paper. For other values, one
  should add $5\log h$.} The statistical errors in measuring void
statistics are dominated by the finite number of independent volumes
in the simulation (or survey), once the voids become large. The $(400$
\hmpc$)^3$ simulation volume is comparable to that of the
volume-limited sample of bright $M_r<-21$ galaxies expected for the
full SDSS. Therefore, the statistical errors in our calculations are
comparable to those expected from the SDSS analysis. The volume of the
full SDSS $M_r < -19$ sample will be approximately the same as one
octant of our simulation, and our errors are therefore smaller by
$\sim \sqrt{8}$. In practice, the observational errors will be
somewhat larger than the error in periodic cubes of the same volume
because edges and ``holes'' in the survey region mean that one cannot
use the full volume for void measurements.

To determine HOD parameters, we use the measurements of \wp\ from
\citet{zehavi04b} for galaxies brighter than $M_r=-21$ and
$M_r=-19$. We choose these two samples because they probe different
regimes of the halo mass function; bright galaxies preferentially
occupy high-mass halos, while the majority of lower luminosity
galaxies occupy lower mass halos (\citealt{zehavi04b}).  We use the
five-parameter HOD model of \cite{zheng04b}, which is flexible enough
to precisely describe the halo occupation functions predicted from
semi-analytic galaxy formation models and hydrodynamic
simulations. The full mean occupation function is the sum of the average
number of central galaxies and satellite galaxies in halos of mass
$M$, i.e. $\navg = \nsat + \ncen$.  For satellite galaxies, the mean
occupation is

\begin{equation}
\label{e.nsat}
\nsat = \left( \frac{M-\mcut}{M_1}\right)^\alpha,
\end{equation}

\noindent
where $M$ is the halo mass, $\mcut$ is a cutoff mass below which a
halo cannot host a satellite galaxy, $M_1+\mcut$ is the mass at which
a halo hosts one satellite on average, and $\alpha$ is the power-law
exponent of the relation, a number usually near one. The distribution
of satellite galaxies about the mean is assumed to be Poisson,
i.e. $\langle N_{\rm sat}(N_{\rm sat}-1)\rangle = \langle N_{\rm
  sat}\rangle^2$. This assumption is well-motivated by both
collisionless and hydrodynamic studies of halo substructure
(\citealt{kravtsov04, zheng04b}), and by observational studies of
galaxies within clusters (\citealt{lin04}).

For central galaxies, the HOD of \cite{zheng04b} contains a soft
transition between one and zero galaxies,

\begin{equation}
\label{e.ncen}
\ncen = \frac{1}{2}\left[ 1+\mbox{erf}\left(\frac{\log M - \log \mmin}{\sigmaM} \right)
	\right],
\end{equation}

\noindent
where $\mbox{erf}(x)$ is the error function, $\mmin$ is characteristic
minimum mass, and $\sigmaM$ is a transition width. As defined in
equation (\ref{e.ncen}), $\mmin$ is the mass at which $\ncen=0.5$. 
All logarithms are base ten. 
As $\sigmaM$ approaches zero, equation (\ref{e.ncen}) approaches a step
function at $\mmin$. 
Since the number of
central galaxies is
assumed to be zero or one by
definition, the distribution of $N_{\rm cen}$ about $\ncen$ is a
nearest-integer distribution (a.k.a. Bernoulli distribution).


\begin{deluxetable*}{ccccccc}
\tablecolumns{7} 
\tablewidth{15pc} 
\tablecaption{HOD Parameters for Models in Figure \ref{wp_fits}}
\tablehead{\colhead{$M_r$} & \colhead{$\s8$} & \colhead{$\slogm$} & \colhead{$\log\mmin$} & \colhead{$\log M_1$} & \colhead{$\alpha$} & \colhead{$\log\mcut$} }
\startdata

-19 & 0.9 & 0.1 & 11.54 & 12.83 & 1.019 & 11.38 \\
-19 & 0.9 & 0.6 & 11.70 & 12.84 & 1.018 & 12.37 \\
-19 & 0.7 & 0.1 & 11.52 & 12.69 & 1.020 & 12.45 \\
-21 & 0.9 & 0.1 & 12.68 & 13.89 & 1.063 & 12.25 \\
-21 & 0.9 & 0.6 & 12.86 & 13.89 & 1.071 & 13.24 \\
-21 & 0.7 & 0.1 & 12.64 & 13.69 & 1.107 & 13.28 \\

\enddata
\tablecomments{All masses are in units of \hmsol. }
\end{deluxetable*}

For full details regarding the calculation of \wp\ for a given
cosmology and HOD, we refer the interested reader to \cite{zheng04a}
and \cite{tinker05a} and the references therein. In the analysis of
this paper, we have used the approach described in \cite{tinker05a}
with one modification; the parameters of the halo bias function listed in
Appendix A ($a=0.707$, $b=0.35$ and $c=0.8$) have been changed to
better match those of the simulation we will be using to calculate
$P_0$: $a$ and $b$ are the same, but $b=0.28$, which lowers the bias
of low-mass halos by about 5\%. (The jackknife errors in the bias of
low-mass halos in the simulation are themselves $\sim 5\%$, so we are
not claiming that this value should be adopted universally.)  The
purpose of our HOD modeling is to find HOD parameters to populate our
simulation. It would make little sense to make mock galaxy populations
that do not match the analytic model or, more importantly, the SDSS
observations.

We assume $\om=0.3$ throughout this paper. For a fixed linear matter
power spectrum, the choice of $\om$ has little influence on the mass
function and clustering of dark matter halos (\citealt{ztwb02}) and
therefore has minimal influence on these calculations. The shape of
$\plin$ is well constrained empirically by the combination of
microwave background anisotropies and large-scale galaxy clustering
measurements (e.g., \citealt{percival02, spergel03, tegmark04}). On a
more practical level, we are required to match the linear matter power
spectrum used to create the initial conditions of the simulation, so
all calculations use the \cmbfast\ power spectrum with the parameters
listed in \S 2.1. This $\plin$ is in good agreement with the
aforementioned observations, though the recent analysis of
\cite{sanchez05} favors a spectrum with somewhat more large scale power.

For a given galaxy sample, the HOD parameters are determined through
$\chi^2$ minimization using the full covariance matrix estimated through
jackknife resampling of the observational sample (see
\citealt{zehavi04b}). The number of free parameters in equations
(\ref{e.nsat}) and (\ref{e.ncen}) is five, but it is reduced to four by
fixing $\mmin$ in order to match the observed space density of
galaxies. To populate the halos, central galaxies are placed at the
center of mass of each halo, while satellite galaxies are placed
randomly about the center of mass assuming a \cite{nfw97} density
profile appropriate for the halo's mass using the parameterization of
\cite{bullock01}. This technique places the galaxies in a spherically
symmetric distribution about the central galaxy, an assumption which has
negligible importance in the context of this work because we are most
interested in halos that host only central galaxies.


\begin{figure*}
\centerline{\epsfxsize=5.5truein\epsffile{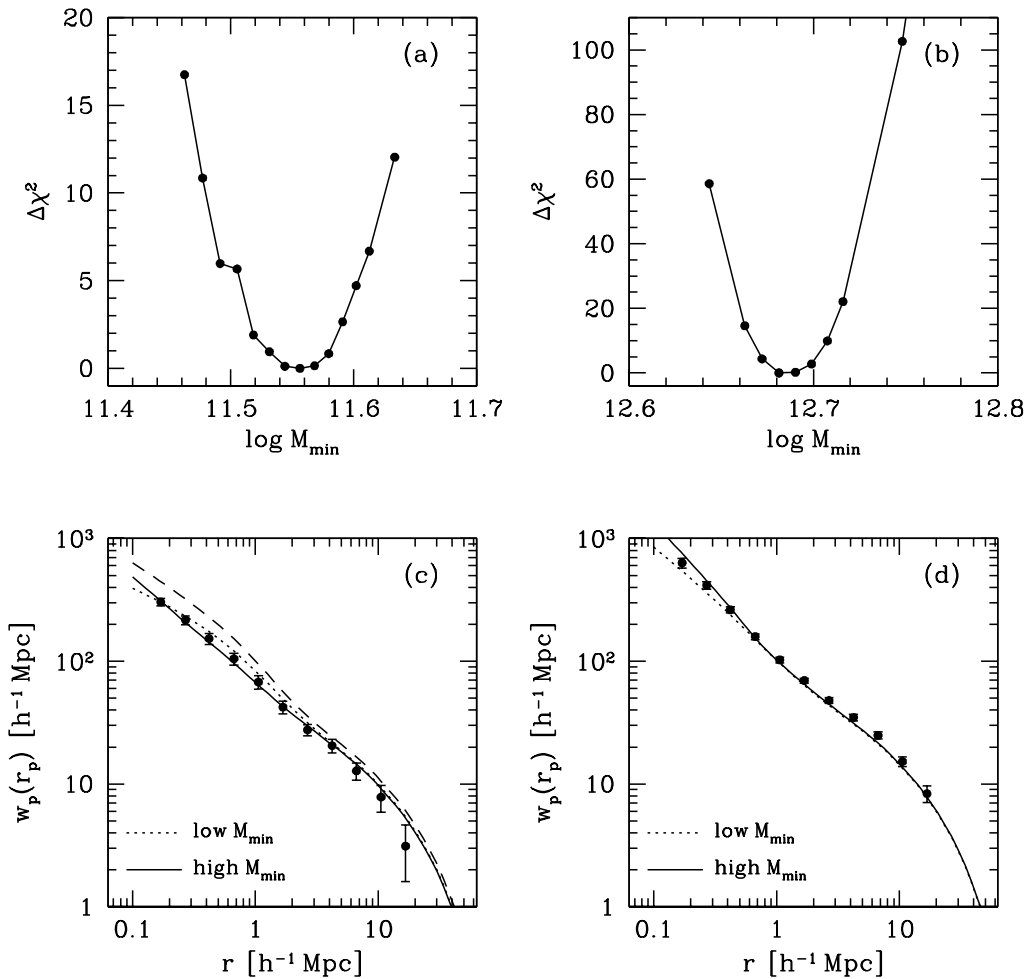}}
\caption{ \label{mmin_wp} Panel (a): $\Delta \chi^2$ of HOD models as
  a function of $\mmin$ with respect to the $M_r<-19$ \wp\ data. All
  models have $\slogm=0.1$, but the parameters that determine $\nsat$
  are left free to match the data. Panel (b): Same as (a), but for
  $M_r<-21$ galaxies. Panel (c): Best-fit HOD models for $M_r<-19$ for
  the highest and lowest values of $\mmin$ with $\Delta \chi^2\approx
  6$ from panel (a). The solid line is $\mmin=4.0\times 10^{11}$
  \hmsol, and the dotted line is $\mmin=3.2\times 10^{11}$ \hmsol. The
  dashed line represents the high-$\mmin$ model with the parameters of
  $\nsat$ taken from the best-fit model from Table 1. Panel (d): Same
  as (c), but for $M_r<-21$. The solid line is $\mmin=5.1\times 10^{12}$
  \hmsol, and the dotted lines is $\mmin=4.7\times 10^{12}$ \hmsol.  }
\end{figure*}


\section{Connecting Voids and \wp\ with the HOD}

\subsection{Variations in the Minimum Mass Scale}

The results of BW demonstrated that the VPF is highly sensitive to the
minimum mass scale, and mostly insensitive to the amplitude and slope of
the occupation function, implying that $\mmin$ can be constrained by
measurements of void statistics. We test this implication with two
significant improvements over the analysis of BW: our central-satellite
approach to the occupation function, as described in \S 2, and a full
statistical comparison of a given HOD model with \wp\ data.

To quantify the range of $\mmin$ allowed by the observed \wp, we first
determine the best-fit values of the HOD parameters, including $\mmin$, for
both the $M_r<-19$ and $M_r<-21$ samples with $\slogm$ fixed at
0.1. This value of $\slogm$ creates an occupation function with a fairly
sharp central cutoff, but the strength of the inferred constraints on
$\mmin$ will be consistent for any reasonable value of $\slogm$. We keep
$\slogm$ fixed to isolate the effect of changing the minimum mass scale
only, without any change in the shape of the central cutoff. The two
parameters that control $\ncen$ have quite different physical
interpretations, so we examine their effect on \wp\ and void
statistics separately. The best-fit values of the HOD parameters for
$\slogm=0.1$ are listed in Table 1. For $M_r<-19$ galaxies, we find
$\mmin=3.60\times 10^{11}$ \hmsol, while for the brighter $M_r<-21$
galaxies, we find $\mmin=4.78\times 10^{12}$ \hmsol. For all
calculations, we assume $\s8=0.9$.

Figure \ref{mmin_wp} demonstrates that the allowed range of $\mmin$ is
quite narrow. Panels (a) and (b) plot $\Delta \chi^2$ for models in
which $\mmin$ is varied around the minimum value for the faint and
bright galaxy samples, respectively. As $\mmin$ is varied, the best-fit
HOD parameters are determined once again by $\chi^2$ minimization, but
with $\mmin$ and $\slogm$ held constant and $M_1$ fixed by the space
density. This leave two free parameters, $\alpha$ and $\mcut$, for the
$\chi^2$ analysis.  For $M_r<-19$ galaxies, $\Delta\log\mmin = \pm
0.05$ produces $\Delta\chi^2 = 6$. The constraints for the $M_r<-21$
sample are even narrower, with $\Delta\log \mmin=\pm 0.015$ producing
$\Delta\chi^2 = 6$. The change in $\chi^2$ is symmetric about the
minimum for both magnitude thresholds. The tighter constraints on
$\mmin$ for $M_r<-21$ are mostly a result of smaller observational
errors for this sample.

The tight constraints on $\mmin$ are a result of the information
\wp\ reveals about the fraction of galaxies that are satellites. The
fraction of satellites $\fsat$ influences the shape and amplitude of
the correlation function at small scales, where pairs from within a
single halo dominate. Once the satellite fraction is known, the value
of $\mmin$ (for a given $\slogm$) is determined by the number density
constraint. Figures \ref{mmin_wp}c and \ref{mmin_wp}d plot the model
\wp\ functions for the two values of $\mmin$ that yield $\Delta
\chi^2=6$ in the upper panels ($3.2\times 10^{11}$ and $4.0\times
10^{11}$ \hmsol\ for $M_r<-19$, and $4.7\times 10^{12}$ and $5.1\times
10^{12}$ \hmsol\ for $M_r<-21$). HOD calculations for the high and low
$\mmin$ are shown with the solid and dotted lines, respectively, while
points with error bars are the SDSS data. As $\mmin$ increases,
$\fsat$ increases in order to preserve the space density of galaxies,
steepening and amplifying the one-halo term of the correlation
function. This behavior is most easily seen in the $M_r<-21$ models in
panel (d). The model \wp\ curves are consistent with each other and
with the data until $r\lesssim 1$ \hmpc, which represents the
transition scale between the one-halo term and the two-halo
term. Below this scale the curves diverge, with the high-$\mmin$ curve
above the low-$\mmin$ curve. 

The behavior for the $M_r<-19$ models in panel (c) is a bit different:
the high-$\mmin$ model has a steeper slope in the one-halo regime, but
its amplitude is lower except at very small scales. In this case, the
adjustment of $\alpha$ and $\mcut$ after changing $\mmin$ provides
enough freedom to lower the one-halo contribution to \wp. If we keep
these parameters fixed in addition to $\slogm$ and the (changed) value
of $\mmin$, we obtain the dashed curve in Figure 1c, which lies well
above the data all scales.


\begin{figure*}
\centerline{\epsfxsize=5.5truein\epsffile{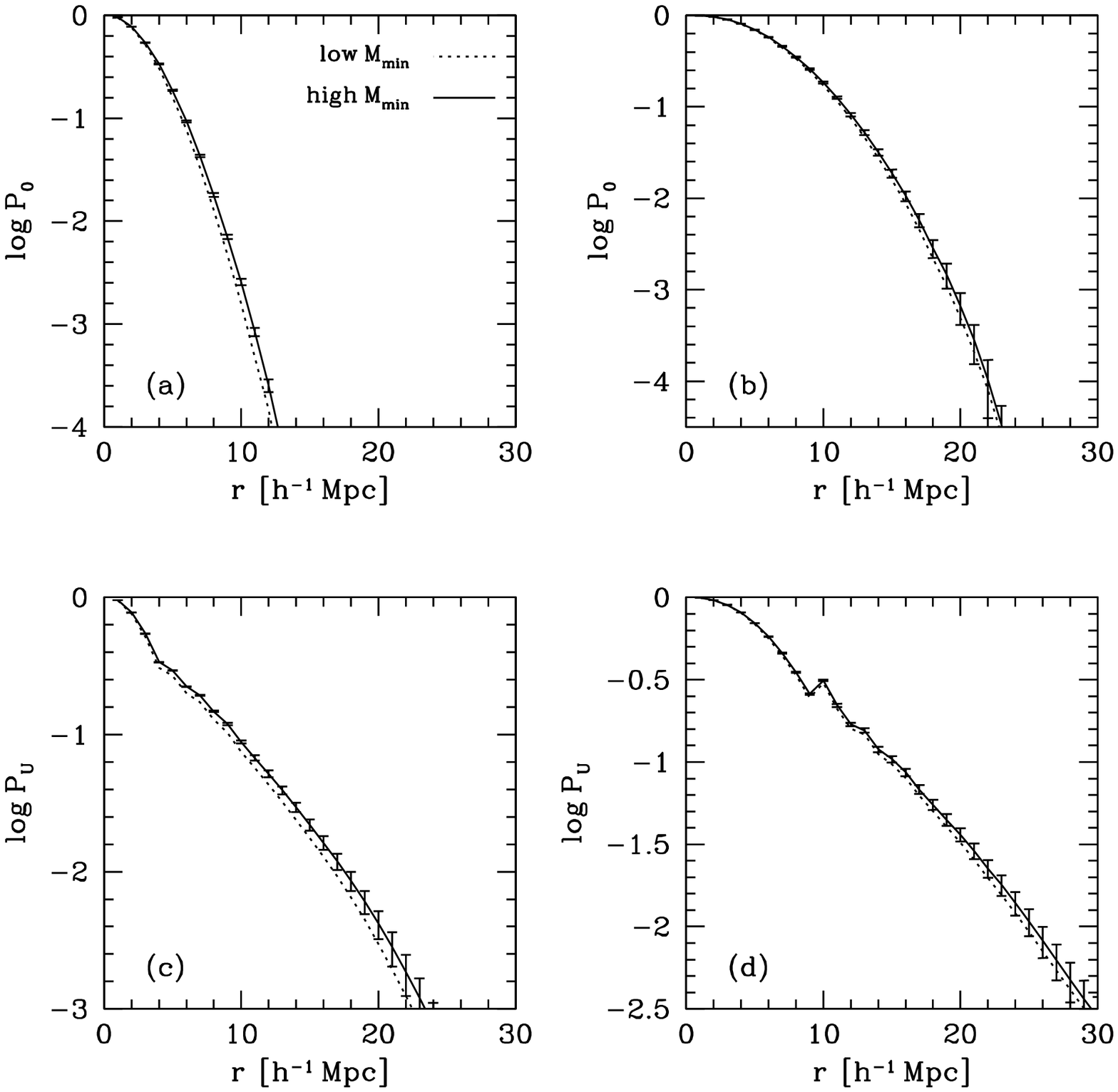}}
\caption{ \label{mmin_vpf} Panel (a): Void probability function (VPF)
  for the high- and low-$\mmin$ models of Figure \ref{mmin_wp}c for
  the $M_r<-19$ galaxy sample. Panel (b):
  Same as (a), but for the $M_r<-21$ sample from Figure \ref{mmin_wp}d. 
  Panel (c): Underdense
  probability distribution (UPF) for the high- and low-$\mmin$ models
  of the $M_r<-19$ sample. Panel (d): Same as (c), but for $M_r<-21$
  galaxies.  }
\end{figure*}

Figure \ref{mmin_vpf} shows the void statistics for the high- and
low-$\mmin$ models of Figure \ref{mmin_wp}. The VPF is calculated for
each galaxy distribution by randomly placing $10^6$ spheres of radius
$r$ within the box and counting the fraction that are empty. The UPF
is calculated in the same manner as the VPF, but each sphere which has
a mean interior density less than 0.2 times the mean galaxy density is
counted. The error at each $r$ is determined from the jackknife
method, dividing the simulation volume into octants. Alternatively,
\cite{hoyle04} estimate observational errors from the error in the
mean of void counts expected if these follow a binomial distribution:

\begin{equation}
\label{e.vpf_error}
\sigma (P_0) = \frac{ (P_0 - P_0^2)^{1/2}}{N_{\rm ind.}^{1/2}},
\end{equation}

\noindent
where $N_{\rm ind.} = 3 R_{\rm box}^3/(4\pi r^3)$ is the number of
independent volumes in the box at each $r$ for which $P_0$ is
calculated. For the VPF, the jackknife errors are consistent with
those determined from equation (\ref{e.vpf_error}) at radii where
$P_0(r)\approx 0.01$, while at larger radii the jackknife errors are
smaller by a factor of two or more. At smaller radii, the jackknife
errors are significantly larger, but the percentage errors are still
less that 1\%. All calculations are done in redshift space using the
distant observer approximation, allowing us to use one axis of the box
as the observer's line of sight and impose periodic boundaries for our
galaxies shifted past the edge of the cube. We find little difference
between the void statistics calculated in and real and redshift space,
as also noted by \cite{little94}.


\begin{figure*}
\centerline{\epsfxsize=5.5truein\epsffile{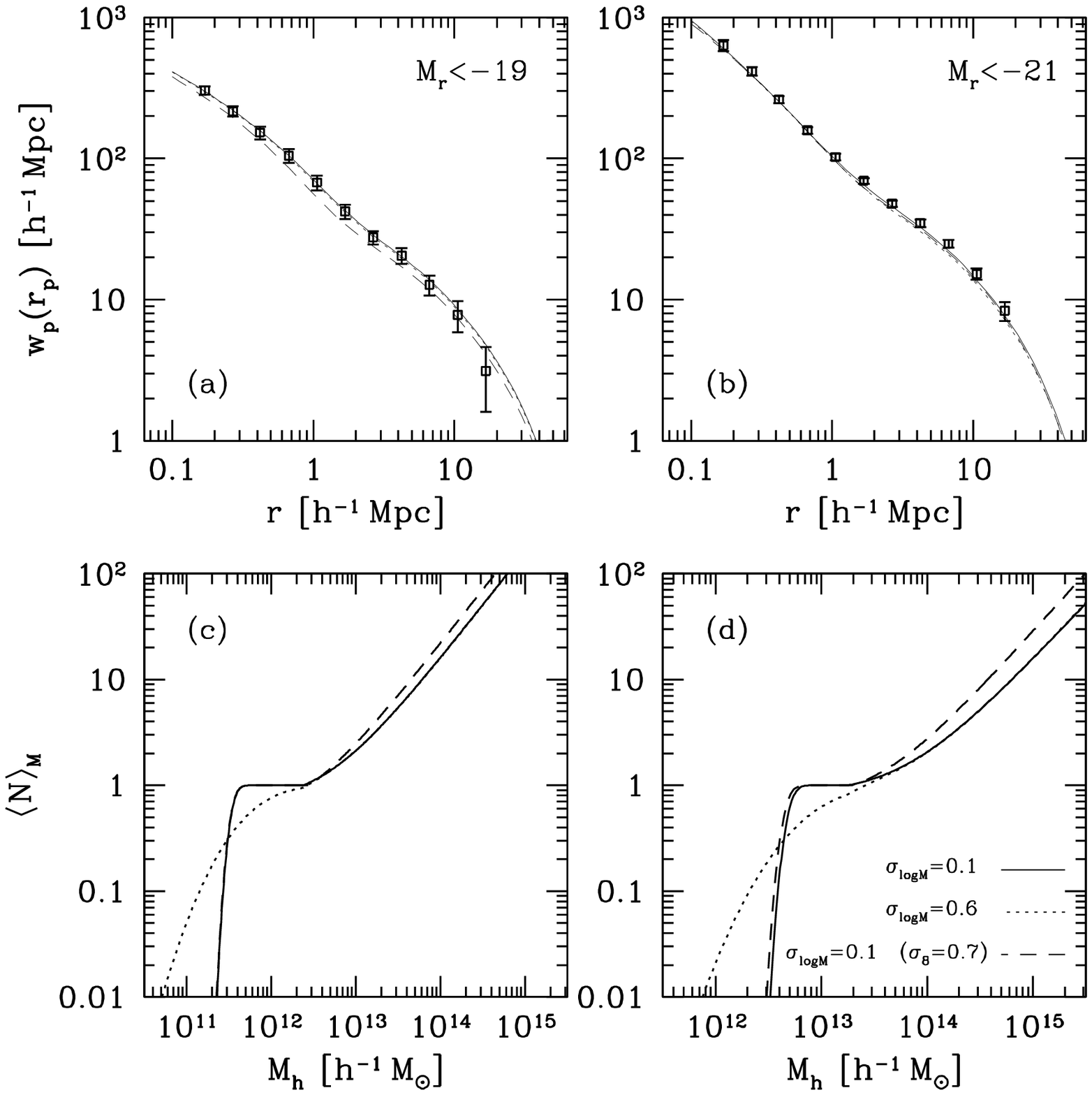}}
\caption{ \label{wp_fits} HOD fits to SDSS samples for $M_r<-19$ (left
  panels) and $M_r<-21$ (right panels). Panels (a) and (b): Projected
  correlation functions. Points with error bars are SDSS data. Lines
  represent HOD models (see legend in panel d). Panels (c) and (d):
  Halo occupation functions derived from the data. The $x$-axis is
  halo mass and the $y$-axis is the average number of galaxies. }
\end{figure*}

For the high space density, $M_r<-19$ sample, the predicted VPF falls
rapidly, dropping to $P_0(r) = 10^{-3}$ at $r=11$ \hmpc\ (Figure
2a). As expected, the low-$\mmin$ model has a lower VPF at all radii,
but because the range in $\mmin$ allowed by the \wp\ and $\ngavg$
constraints is so small, the difference between the low- and
high-$\mmin$ models is small. Since our errors bars for this simulated
sample should be smaller than those for the full SDSS, the difference
is likely to be indiscernible in practice. Note that errors at
different $r$ are highly correlated, as any large void would also
contain smaller empty spheres, so the points do not scatter about a
smooth curve by the amount of their error bars. For the $M_r<-21$
sample, voids are larger, with $P_0(r)=10^{-3}$ at $r=20$
\hmpc\ (Figure 2b). The tight constraints on $\mmin$ for this sample
lead to a very small difference between the low- and high-$\mmin$
VPFs, again smaller than the expected SDSS errors.

The UPFs for the faint and bright samples are shown in Figures
\ref{mmin_vpf}c and \ref{mmin_vpf}d, respectively. It is more likely
to find a region with low density contrast than a completely empty
region, so $P_U \ge P_0$ at all $r$. The UPF is also not as steep a
function of $r$, since the $n_g\le 0.2\ngavg$ threshold allows a
larger number of galaxies in larger radius spheres. The results for
the UPF parallel the results of the VPF; there is a small difference
in the low- and high-$\mmin$ models, with high-$\mmin$ producing
higher $P_U$, but the difference between the models is only marginally
significant.

BW reached different conclusions about the sensitivity of the VPF to
$\mmin$ because they did not impose \wp\ as a constraint. They varied
$\mmin$ by a factor of 8, producing radical changes in the VPF, but
also changing the correlation function. The VPF {\it is} sensitive to
$\mmin$, but Figures 1 and 2 show that \wp\ constrains $\mmin$ tightly
enough that void statistics cannot add much further leverage. This
result was also found by \cite{conroy05} in their analysis of
voids in the DEEP2 and SDSS surveys. The constraints on $\mmin$ in
Figure \ref{mmin_wp} will become tighter as the full sample of SDSS
galaxies is analyzed, significantly reducing the errors on \wp.


\begin{deluxetable}{ccc}
\tablecolumns{3} 
\tablewidth{13pc} 
\tablecaption{Predicted void Statistics for $M_r<-19$ Galaxies}
\tablehead{ \colhead{$r$ [\hmpc ]} & \colhead{$P_0(r)$} & \colhead{$P_U(r)$} }

\startdata

1 & $( 9.50 \pm 0.01 ) \times 10^{-1} $ & $( 9.50 \pm 0.00 ) \times 10^{-1} $ \\ 
2 & $( 7.62 \pm 0.02 ) \times 10^{-1} $ & $( 7.61 \pm 0.02 ) \times 10^{-1} $ \\ 
3 & $( 5.21 \pm 0.02 ) \times 10^{-1} $ & $( 5.21 \pm 0.03 ) \times 10^{-1} $ \\ 
4 & $( 3.12 \pm 0.03 ) \times 10^{-1} $ & $( 3.12 \pm 0.03 ) \times 10^{-1} $ \\ 
5 & $( 1.67 \pm 0.02 ) \times 10^{-1} $ & $( 2.75 \pm 0.03 ) \times 10^{-1} $ \\ 
6 & $( 7.95 \pm 0.19 ) \times 10^{-2} $ & $( 2.07 \pm 0.03 ) \times 10^{-1} $ \\ 
7 & $( 3.51 \pm 0.12 ) \times 10^{-2} $ & $( 1.79 \pm 0.03 ) \times 10^{-1} $ \\ 
8 & $( 1.39 \pm 0.07 ) \times 10^{-2} $ & $( 1.35 \pm 0.03 ) \times 10^{-1} $ \\ 
9 & $( 5.10 \pm 0.35 ) \times 10^{-3} $ & $( 1.08 \pm 0.03 ) \times 10^{-1} $ \\ 
10 & $( 1.78 \pm 0.20 ) \times 10^{-3} $ & $( 7.92 \pm 0.33 ) \times 10^{-2} $ \\ 
11 & $( 5.44 \pm 0.77 ) \times 10^{-4} $ & $( 6.14 \pm 0.31 ) \times 10^{-2} $ \\ 
12 & $( 1.51 \pm 0.36 ) \times 10^{-4} $ & $( 4.68 \pm 0.29 ) \times 10^{-2} $ \\ 
13 & $( 2.60 \pm 0.95 ) \times 10^{-5} $ & $( 3.53 \pm 0.26 ) \times 10^{-2} $ \\ 
14 & $( 1.00 \pm 1.00 ) \times 10^{-7} $ & $( 2.64 \pm 0.24 ) \times 10^{-2} $ \\ 
15 &  --- & $( 1.96 \pm 0.21 ) \times 10^{-2} $ \\ 
16 &  --- & $( 1.44 \pm 0.19 ) \times 10^{-2} $ \\ 
17 &  --- & $( 1.05 \pm 0.16 ) \times 10^{-2} $ \\ 
18 &  --- & $( 7.49 \pm 1.38 ) \times 10^{-3} $ \\ 
19 &  --- & $( 5.24 \pm 1.17 ) \times 10^{-3} $ \\ 
20 &  --- & $( 3.56 \pm 0.97 ) \times 10^{-3} $ \\ 
21 &  --- & $( 2.38 \pm 0.78 ) \times 10^{-3} $ \\ 
22 &  --- & $( 1.57 \pm 0.62 ) \times 10^{-3} $ \\ 
23 &  --- & $( 1.01 \pm 0.47 ) \times 10^{-3} $ \\ 
24 &  --- & $( 6.31 \pm 3.51 ) \times 10^{-4} $ \\ 
25 &  --- & $( 3.85 \pm 2.53 ) \times 10^{-4} $ \\

\enddata
\end{deluxetable}

\subsection{Variations in $\s8$ and the Central Cutoff}

The shape of the central galaxy cutoff reflects the scatter between
galaxy properties and host halo mass. A soft cutoff corresponds to
significant scatter, allowing some halos with $M\ll \mmin$ to host a
galaxy within the sample and some halos with $M>\mmin$ to be scattered
out of the sample. In the context of our luminosity-threshold samples
the scatter is between luminosity and mass, but it can be reflected in
any quantity used to define the sample, such as surface brightness,
color, or nuclear activity.

The shape of the central cutoff may alter void statistics because the
halos that outline the voids are low mass. Numerical simulations of
void regions show that the maximum halo mass within voids increases
from the center of the void to its outer edge
(\citealt{gottlober03}). Allowing some galaxies to occupy halos below
$\mmin$ could effectively ``shrink'' the voids. To quantify the effect
of the shape of $\ncen$, we fix $\sigmaM$ at six values ranging evenly
from 0.1 to 0.6. At $\sigmaM=0.1$, $\ncen$ is nearly a step function
at $\mmin$, while for $\sigmaM=0.6$ the probability of finding a
galaxy in a halo of mass $\mmin/10$ is still $\sim 1\%$, which can be
significant given the steepness of the halo mass function. For each
value of $\slogm$, we find the best-fit values of the four remaining
HOD parameters by $\chi^2$ minimization.

We also investigate the influence of the dark matter clustering
amplitude by considering a model with $\slogm=0.1$ but
$\s8=0.7$. Power spectra with lower amplitudes produce smaller voids
in the matter distribution, but the effect on galaxy voids is not
immediately clear. A model with lower $\s8$ must have higher galaxy
bias to match \wp. This requires shifting more galaxies into higher
mass halos, and thus out of low-density regions.

Figure \ref{wp_fits}a plots \wp\ for the data for $M_r<-19$ and for
three of the HOD models: $(\s8,\slogm) = (0.9,0.1)$, $(0.9,0.6)$, and
$(0.7,0.1)$. The two models with higher $\s8$ are excellent fits to
the observations and are nearly indistinguishable both visually and in
$\chi^2$. Panel (c) plots $\navg$ against $M$ for these three
models. Increasing $\slogm$ produces virtually identical correlation
functions because the bias function of halos below $\sim 10^{12}$
\hmsol, where the differences in $\navg$ begin, is relatively
flat. Redistributing galaxies from halos of $2\times 10^{12}$
\hmsol\ into lower mass halos near $10^{11}$ \hmsol\ does not
significantly change the large scale galaxy bias factor (0.94 and 0.93
for $\slogm=0.1$, 0.6). The large-scale clustering is thus
unchanged. The differences in $\nsat$ are minimal, so the one-halo
term is also unaltered. The only change in \wp\ from $\slogm=0.1$
to 0.6 is in the two-halo term at small scales, $r\lesssim 0.8$ \hmpc,
but at these scales \wp\ is dominated by the one-halo term and the
changes cannot be distinguished. The fit for the $\s8=0.7$ model does
not match the data as well and has a $\Delta \chi^2$ of approximately
five relative to the $\s8=0.9$ fits. The low amplitude of matter
clustering for this model can only be partially overcome by creating
an occupation function that has more galaxies in high-mass
halos. Changes to $\slogm$ make little difference in the agreement.

Figure \ref{wp_fits}b shows the HOD fits to the $M_r<-21$ sample for
the same three models. All three appear nearly equivalent, but
statistically, the $(\s8,\slogm) = (0.9,0.1)$ model has $\chi^2=5.5$,
while the $(0.9,0.6)$ model has $\chi^2=11.4$ and $(0.7,0.1)$ has
$\chi^2=10.0$.  For both values of $\s8$, the $\chi^2$ of the HOD fit
increases monotonically with $\sigmaM$. In contrast to the $M_r<-19$
sample, moving galaxies from $10^{13}$ \hmsol\ halos into smaller ones
has a noticeable effect on the average halo bias of the sample (the
galaxy bias is 1.20 and 1.15 for $\slogm=0.1$ and $0.6$
respectively). The subsequent HOD model is less able to match the
observed amplitude of galaxy clustering.


\begin{figure}[t]
\centerline{\epsfxsize=3.2truein\epsffile{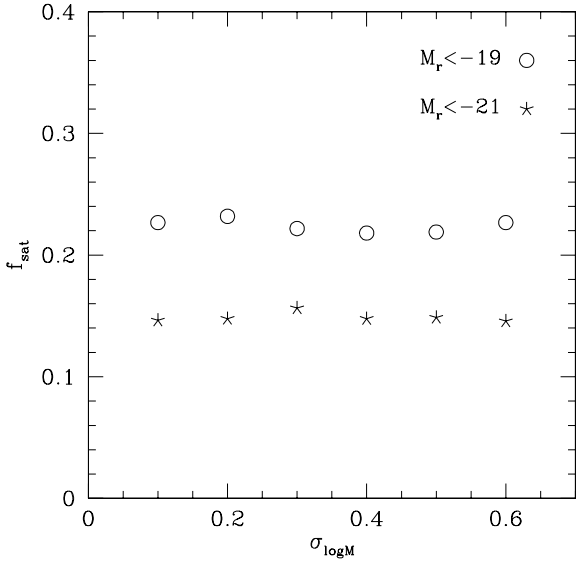}}
\caption{ \label{fsat} Fraction of galaxies that are satellites,
  $\fsat$, as a function of $\slogm$ for both $M_r<-19$ and $M_r<-21$
  galaxies. }
\end{figure}


\begin{figure*}
\centerline{\epsfxsize=5.5truein\epsffile{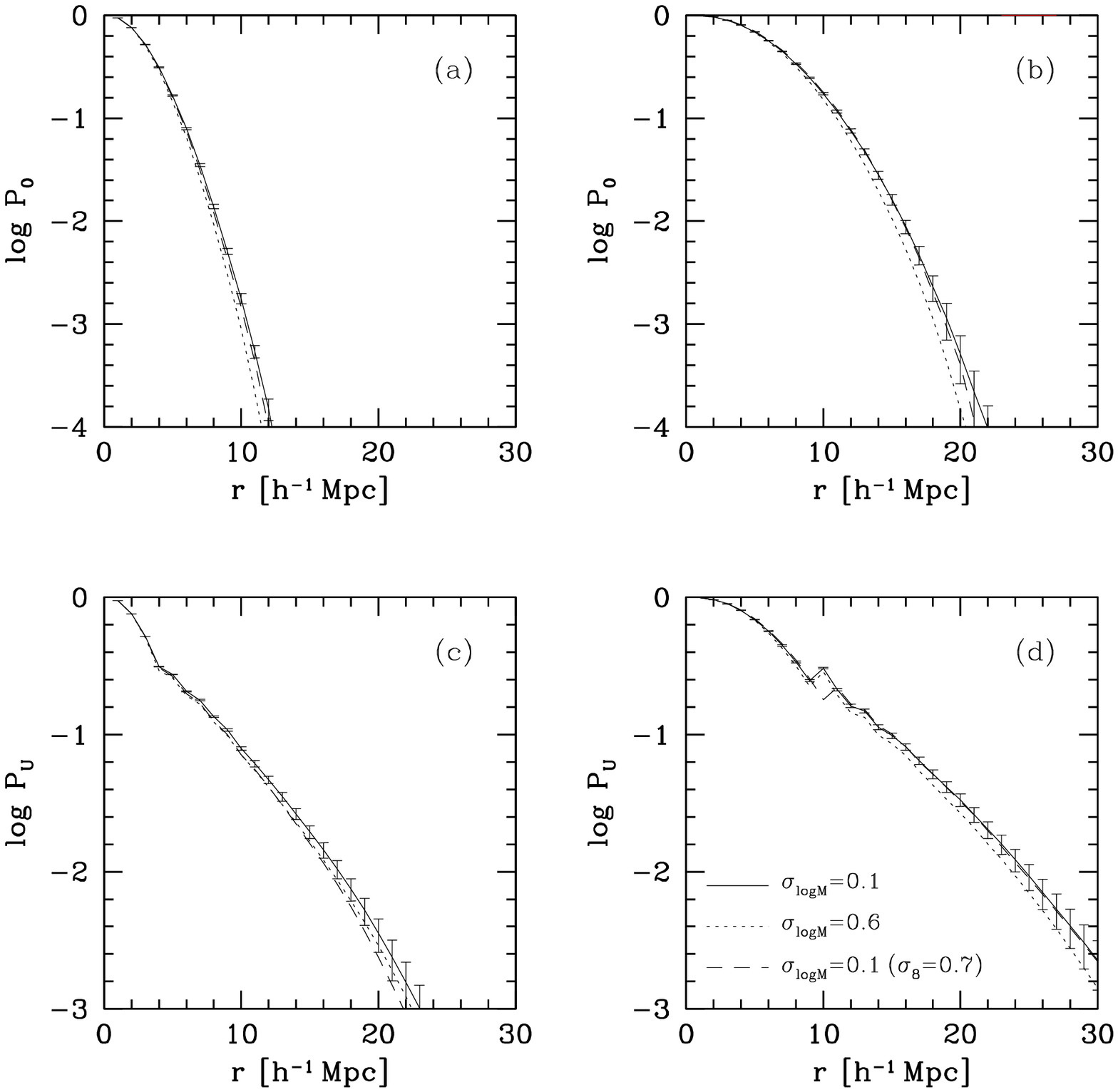}}
\caption{ \label{vstats} Void statistics for the HOD models shown in
  Figure \ref{wp_fits}. Left panels present results for $M_r<-19$, and
  right panels show results for $M_r<-21$. Panels (a) and (b): Void
  probability functions for the three HOD models shown in Figure
  \ref{wp_fits}. Error bars are shown for the $(\s8,\slogm)=(0.9,0.1)$
  models only, but are the same for other models. Panels (c) and (d):
  Underdensity probability functions for the same models. }
\end{figure*}

Figure \ref{fsat} shows the values of $\fsat$ determined from the
best-fit HOD parameters as a function of $\slogm$, for $\s8=0.9$. For
the $M_r<-19$ sample, $\fsat$ for all models is approximately 0.23,
with only modest variation around this mean. For $M_r<-21$ galaxies,
$\fsat\approx 0.15$ for all $\slogm$ values. For higher $\slogm$, the
best-fit $\mmin$ is also higher to compensate for the additional
number of central galaxies occupying $M<\mmin$ halos. But regardless
of the shape of $\ncen$, the correlation function drives the HOD to
the same value of $\fsat$, reaffirming our conclusion from \S 3.1 that
the satellite fraction is a fundamental quantity for determining the
shape and amplitude of the correlation function.

Figure \ref{vstats} plots the void statistics for the same HODs and
luminosity samples presented in Figure \ref{wp_fits}. Panel (a) plots
$P_0(r)$ for the three HOD models of the $-19$ sample. For $\s8=0.9$,
the $\slogm=0.6$ model has smaller voids, as expected. Since $\fsat$
is the same for $\slogm=0.1$ and $0.6$, the two models have equal
number densities of single-galaxy halos, but the $\slogm=0.6$ model
allows some galaxies in lower mass halos that are more likely to be
found in void interiors. However, while the sign of the effect is as
expected, the magnitude is small, and only slightly larger than our
statistical error bars. For example, the radius at which
$P_0(r)=10^{-3}$ changes by less than 1 \hmpc. Results for the UPF
(Figure 5c) and for the $M_r<-21$ sample (Figures 5b and 5d) are similar:
the $\slogm=0.6$ model has smaller voids, but the changes are too
small to be readily detectable.

Dashed curves in each panel in Figure 5 show the $(\s8,\slogm) =
(0.7,0.1)$ model predictions. The increased bias almost exactly
compensates for the lower matter clustering amplitude. There is a
slightly lower UPF for the $M_r<-19$ sample, but this difference is
small, and the other predictions are nearly indistinguishable from
those of the $(\s8,\slogm) = (0.9,0.1)$ model. 

\begin{figure}[t]
\centerline{\epsfxsize=3.2truein\epsffile{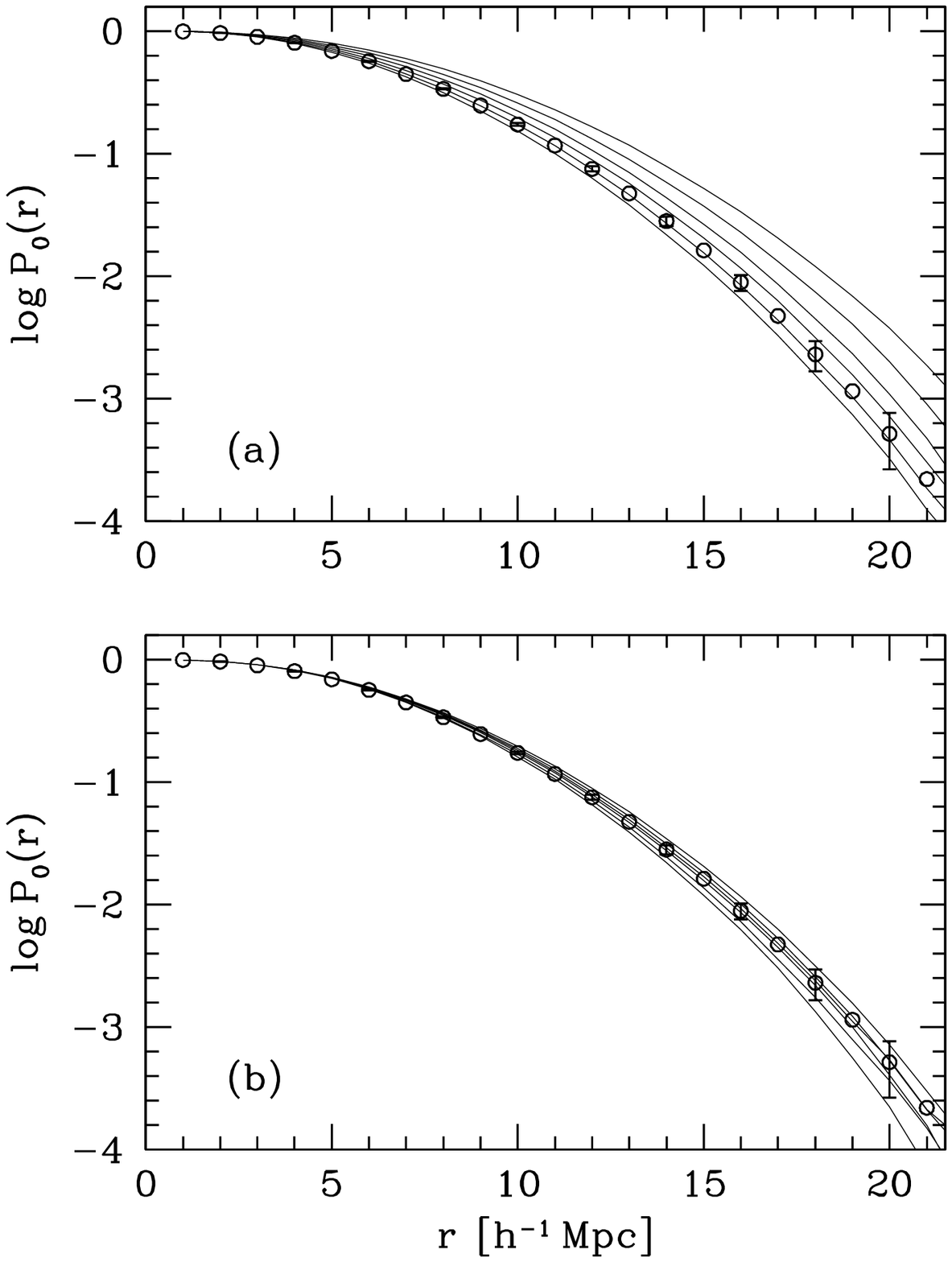}}
\caption{ \label{maxcen_satfrac} Panel (a): VPFs for models with
  varying $\fsat$ but fixed overall space density of galaxies. From
  bottom to top, the 6 solid lines represent $\fsat=0$ to 0.5 in steps
  of 0.1. The open circles represent the $M_r<-21$ results from Figure
  \ref{vpf}b, which have $\fsat=0.15$. Panel (b): VPFs for models with
  varying $\ncen$, but fixed $\fsat=0.2$. From bottom to top, the
  solid lines represent $\ncen=0.5$ to 1.0 in steps of 0.1 (and are
  constant for all halo masses). The open circles are the same as
  panel (a).}
\end{figure}

The insensitivity of void statistics to $\slogm$ implies that
sub-$\mmin$ halos avoid the low density regions almost as strongly as
$\mmin$ halos. The total number density of occupied low mass halos
matters, but it makes little difference whether galaxies populate the
``high end'' of the low mass halo regime or populate it
uniformly. Figure \ref{maxcen_satfrac} demonstrates this point with
two pedagogically illustrative model sequences. In the first, we vary
$\fsat$ while keeping the space density of galaxies fixed to that of
the $M_r<-21$ sample. These models are no longer constrained to match
the correlation function, and the role of $\fsat$ is only to determine
the number density of occupied halos, i.e., $n_h =
(1-\fsat)\ngavg$. For each $\fsat$, $\mmin$ is set by the constraint

\begin{equation}
\label{e.fsat}
n_h = \int_{\mmin}^\infty\,\frac{dn}{dM}\,dM,
\end{equation}

\noindent
and all halos $M\ge\mmin$ are given a single, central galaxy. Figure
\ref{maxcen_satfrac}a plots the VPF for these models with $\fsat=0$ to
0.5 in steps of 0.1. For comparison, the VPF for the $M_r<-21$ sample
with $\slogm=0.1$ is shown with the open circles. The satellite
fraction for this model is 0.15, and its predictions lie in between
those of the $\fsat=0.1$ and $\fsat=0.2$ models. The VPF monotonically
increases with increasing $\fsat$ because the number density of
occupied halos decreases. To demonstrate the relative unimportance of
spreading galaxies over a range of low mass halos, we have created a
model sequence in which $\fsat$ is held constant but $\ncen$ is set to
be a constant number less than unity at all masses. In these models,
$\mmin$ is set by the constraint

\begin{equation}
\label{e.fsat}
n_h = \int_{\mmin}^\infty\,\ncen\,\frac{dn}{dM}\,dM,
\end{equation}

\noindent
using the nearest integer distribution to place central galaxies in
halos. Figure \ref{maxcen_satfrac}b shows the VPF for models in which
$\fsat=0.2$ and $\ncen=1.0$ to 0.5 in steps of 0.1. In these models the
number density of occupied halos is constant, but as $\ncen$ decreases
the average mass of occupied halos decreases. The results in Figure
\ref{maxcen_satfrac}b show a trend of lower VPF for lower $\ncen$, but
the amplitude of the effect is small compared to that of varying $\fsat$.

These results demonstrate that models that are constrained to match
\wp\ offer a remarkably robust prediction of the VPF and UPF,
regardless of the form of the HOD or the value of $\s8$ (at least in
the range $0.7-0.9$ considered here). We list the values of $P_0$ and
$P_U$ determined for the $(\s8,\slogm)=(0.9,0.1)$ models in Tables 2
and 3 for $M_r<-19$ and $M_r<-21$, respectively.


\begin{deluxetable}{ccc}
\tablecolumns{3} 
\tablewidth{13pc} 
\tablecaption{Predicted void Statistics for $M_r<-21$ Galaxies}
\tablehead{
\colhead{$r$ [\hmpc ]} & \colhead{$P_0(r)$} & \colhead{$P_U(r)$}  }
\startdata

1 & $( 9.95 \pm 0.01 ) \times 10^{-1} $ & $( 9.95 \pm 0.01 ) \times 10^{-1} $ \\ 
2 & $( 9.65 \pm 0.01 ) \times 10^{-1} $ & $( 9.65 \pm 0.01 ) \times 10^{-1} $ \\ 
3 & $( 8.99 \pm 0.02 ) \times 10^{-1} $ & $( 8.99 \pm 0.02 ) \times 10^{-1} $ \\ 
4 & $( 8.05 \pm 0.03 ) \times 10^{-1} $ & $( 8.05 \pm 0.03 ) \times 10^{-1} $ \\ 
5 & $( 6.90 \pm 0.03 ) \times 10^{-1} $ & $( 6.90 \pm 0.03 ) \times 10^{-1} $ \\ 
6 & $( 5.68 \pm 0.04 ) \times 10^{-1} $ & $( 5.68 \pm 0.04 ) \times 10^{-1} $ \\ 
7 & $( 4.47 \pm 0.04 ) \times 10^{-1} $ & $( 4.47 \pm 0.04 ) \times 10^{-1} $ \\ 
8 & $( 3.39 \pm 0.05 ) \times 10^{-1} $ & $( 3.39 \pm 0.05 ) \times 10^{-1} $ \\ 
9 & $( 2.47 \pm 0.04 ) \times 10^{-1} $ & $( 2.47 \pm 0.04 ) \times 10^{-1} $ \\ 
10 & $( 1.73 \pm 0.04 ) \times 10^{-1} $ & $( 3.05 \pm 0.04 ) \times 10^{-1} $ \\ 
11 & $( 1.17 \pm 0.04 ) \times 10^{-1} $ & $( 2.13 \pm 0.05 ) \times 10^{-1} $ \\ 
12 & $( 7.51 \pm 0.37 ) \times 10^{-2} $ & $( 1.62 \pm 0.04 ) \times 10^{-1} $ \\ 
13 & $( 4.73 \pm 0.30 ) \times 10^{-2} $ & $( 1.49 \pm 0.05 ) \times 10^{-1} $ \\ 
14 & $( 2.81 \pm 0.25 ) \times 10^{-2} $ & $( 1.14 \pm 0.04 ) \times 10^{-1} $ \\ 
15 & $( 1.62 \pm 0.19 ) \times 10^{-2} $ & $( 9.82 \pm 0.45 ) \times 10^{-2} $ \\ 
16 & $( 8.89 \pm 1.31 ) \times 10^{-3} $ & $( 8.15 \pm 0.43 ) \times 10^{-2} $ \\ 
17 & $( 4.72 \pm 0.96 ) \times 10^{-3} $ & $( 6.45 \pm 0.41 ) \times 10^{-2} $ \\ 
18 & $( 2.30 \pm 0.64 ) \times 10^{-3} $ & $( 5.16 \pm 0.39 ) \times 10^{-2} $ \\ 
19 & $( 1.15 \pm 0.45 ) \times 10^{-3} $ & $( 4.16 \pm 0.37 ) \times 10^{-2} $ \\ 
20 & $( 5.15 \pm 2.51 ) \times 10^{-4} $ & $( 3.34 \pm 0.34 ) \times 10^{-2} $ \\ 
21 & $( 2.19 \pm 1.30 ) \times 10^{-4} $ & $( 2.62 \pm 0.31 ) \times 10^{-2} $ \\ 
22 & $( 9.50 \pm 6.54 ) \times 10^{-5} $ & $( 2.03 \pm 0.28 ) \times 10^{-2} $ \\ 
23 & $( 3.10 \pm 2.05 ) \times 10^{-5} $ & $( 1.59 \pm 0.25 ) \times 10^{-2} $ \\ 
24 & $( 1.20 \pm 1.20 ) \times 10^{-5} $ & $( 1.23 \pm 0.23 ) \times 10^{-2} $ \\ 
25 & $( 3.00 \pm 3.00 ) \times 10^{-6} $ & $( 9.31 \pm 2.01 ) \times 10^{-3} $ \\ 

\enddata
\end{deluxetable}


\section{Density Dependence of the HOD}

Motivated by the points discussed in \S 1, we now consider models in
which the HOD---specifically the minimum mass scale for central
galaxies---changes systematically with environment. To the extend that
a galaxy sample is defined by an observable property that correlates
strongly with halo formation time, the dependence of formation time on
large scale overdensity (\citealt{gao05, harker05, wechsler05}) will drive such
variations in the HOD. The strength of this dependence increases
rapidly toward lower halo masses, so it is interesting to examine
environmental effects for both faint and bright galaxy samples.

\begin{figure}[t]
\centerline{\epsfxsize=3.2truein\epsffile{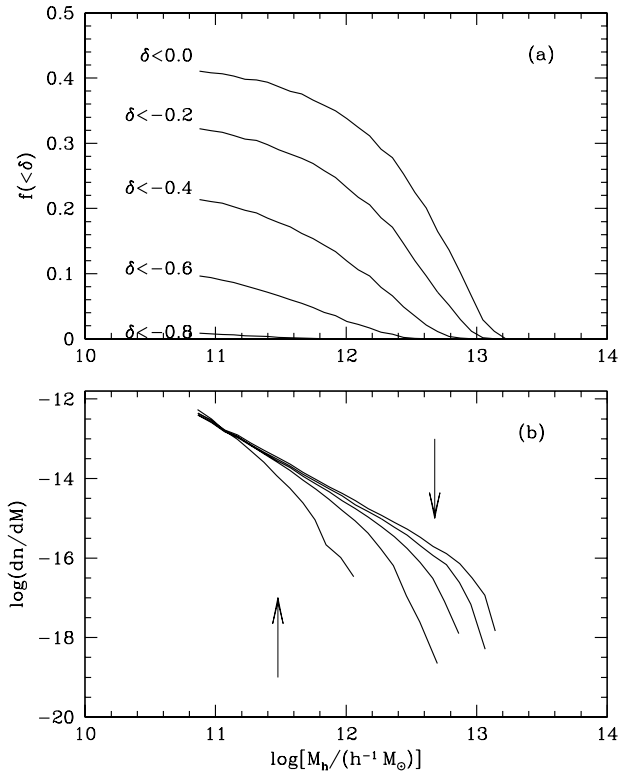}}
\caption{ \label{f_delta} Panel (a): Fraction of halos that lie in
  regions below a given matter density as a function of halo mass. The
  top line represents the fraction of halos that lie in underdense
  ($\delta<0$) regions as a function of halo mass. Subsequent lines
  present results for lower density contrasts as labeled in the
  panel. Panel (b): Mass functions of halos in low-density
  environments. Solid lines show the mass functions for halos in
  $\delta<-0.8$ regions (far left) to $\delta<0$ (far right),
  corresponding to the bottom and top lines in panel (a)
  respectively. The upward pointing arrow marks the location of
  $\mmin$ for the $M_r<-19$ sample, and the downward arrow indicates
  $\mmin$ for $M_r<-21$ galaxies (for $\slogm=0.1$). For all density
  calculations, the smoothing radius is 5 \hmpc.}
\end{figure}

Even in the deepest voids in the dark matter, halos still form
(\citealt{gottlober03}). Figure \ref{f_delta}a shows the fraction of
halos below a given overdensity as a function of halo mass. For each
halo in the simulation, the surrounding dark matter density is
calculated in a top-hat sphere 5 \hmpc\ in radius. Although the choice
of smoothing radius is to some degree arbitrary, we adopt 5
\hmpc\ over a larger radius because environmental dependence in
low-density regions may be sensitive not just to whether the halo
resides in a void, but also to {\it where} in the void, and the local
matter density increases closer to the edge.  At $10^{11}$
\hmsol\ approximately 40\% of the halos are in underdense regions, and
$\sim 10\%$ are in regions with $\delta \le -0.4$. These fractions
monotonically decrease with increasing halo mass, until the fraction
of halos in underdense environments is nearly zero for all halos above
$10^{13}$ \hmsol. Figure \ref{f_delta}b plots the mass function for
the halos in different environments. From left to right, the solid
lines represent mass functions for halos in regions with $\delta<-0.8$
to $\delta<0$ in steps of $0.2$. Each mass function can be
approximated by a power-law at low masses with an exponential cutoff
at high masses (\citealt{ps}). The change from power-law to
exponential occurs at a mass scale denoted $\mstar$.  The value of
$\mstar$ for each mass function plotted in panel (b) corresponds to
the mass at which the curves in panel (a) turn over and begin to
approach zero. The arrows indicate $\mmin$ for $M_r<-19$ (upward
arrow) and $M_r<-21$ (downward arrow), indicating that bright galaxies
reside in regions with matter densities $\delta\gtrsim -0.2$, but that
faint galaxies reside in nearly all environments.


Our model for the density dependence of halo occupation is simple:
below a given density threshold $\dc$, the minimum mass that can host
a galaxy (in the galaxy class being modeled) increases by a factor
$\fmin$. For $M_r<-19$, we will examine models in which $\dc$ ranges
from $-0.8$ to $-0.2$ and $\fmin=2$, 4, and $\infty$ (i.e., galaxy
formation is halted below $\dc$ for this class of galaxies). A
physical model would presumably predict a smooth transition in which
the probability of hosting a sample galaxy changes continuously with
local density, but the quantitative effect is most likely degenerate
with our two-parameter model. In our parameterization it is
straightforward to isolate the two characteristics of environmental
dependence: the density at which a change occurs, and magnitude of the
shift in the HOD.

Although the relative number of galaxies in extremely underdense
regions is low, altering the galaxy distribution in these regions does
have some impact on the correlation function. Removing galaxies from
anti-biased halos (and placing them in higher-density environments in
order to keep the total space density of galaxies fixed) increases the
overall bias of the galaxy sample. This increase can be masked to some
degree by changes in other HOD parameters. For each $\dc$ and $\fmin$
we recalculate HOD parameters to obtain a minimum $\chi^2$ fit to the
SDSS \wp\ measurements, but we now use the populated halos of the
N-body simulation to calculate \wp\ rather than the analytic
model. We leave the value of $\slogm$ as a free parameter and start
the minimization at the HOD for the density-independent HOD found in
the previous section for $\slogm=0.1$.


\begin{figure*}
\centerline{\epsfxsize=4.5truein\epsffile{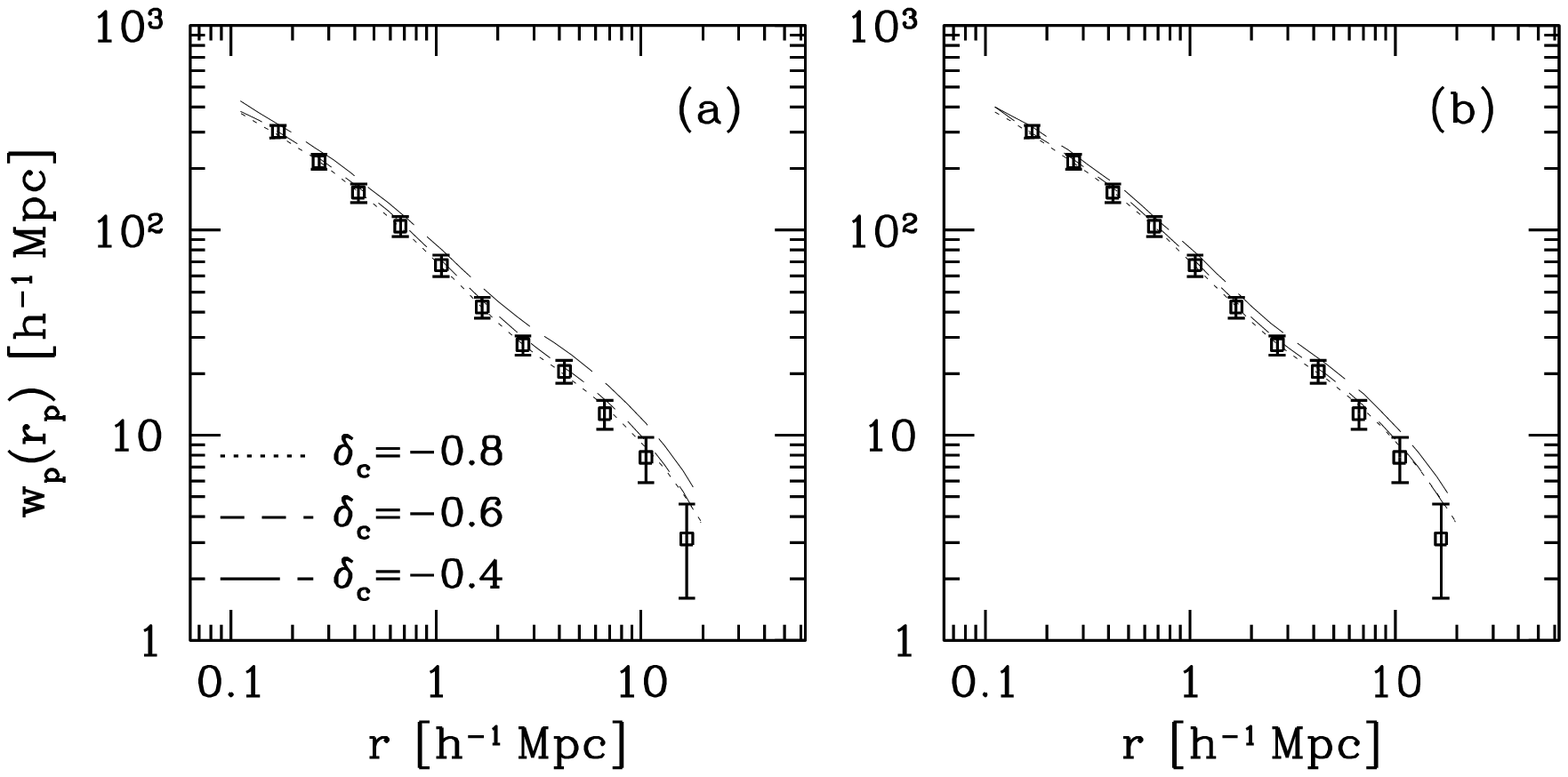}}
\caption{ \label{wp_den} Projected correlation functions for
  density-dependent HOD models of the $M_r<-19$ sample. (a) Models in
  which $M_r<-19$ galaxies are completely suppressed ($\fmin=\infty$)
  below critical densities $\dc = -0.8$, $-0.6$, and $-0.4$. Points
  with error bars are the SDSS data. (b) Models in which the minimum
  host halo mass increases by a factor $\fmin=2$ below the same values
  of $\dc$. }
\end{figure*}

\subsection{Results for $M_r <-19$}

Figure \ref{wp_den} shows several examples of \wp\ for
density-dependent models of the $M_r<-19$ sample. Panel (a) plots the
correlation functions for $\fmin=\infty$ and $\dc=-0.8$, $-0.6$, and
$-0.4$. The SDSS data are also shown for comparison. For $\dc=-0.8$,
the change in the correlation function is negligible. For $\dc=-0.6$,
the exclusion of low-density halos noticeably increases the
large-scale amplitude of \wp. This effect is magnified for the
$\dc=-0.4$ model, and the resulting correlation function is clearly
excluded by the data. The $\Delta \chi^2$ between $\dc=-0.8$ and
$-0.6$ is 1.4, while for $\dc=-0.4$ the $\Delta \chi^2 \approx
10$. Panel (b) in figure \ref{wp_den} shows the results for
$\fmin=2$. This more moderate model of density dependence, where
galaxy formation is suppressed but not eliminated in low density
environments, has a smaller effect on \wp\ at the same value of
$\dc$. 

The effect of density dependence on \wp\ can be interpreted through
the ``voided Poisson'' model of \cite{babul91} and
\cite{babulwhite91}. Removing galaxies from low-density regions makes
the voids larger. Since the galaxy number density is held fixed, a
simple random distribution of the displaced galaxies throughout the
volume not occupied by the voids increases the clustering at
scales less than the characteristic size of the voids. The stronger
the density dependence, the higher the boost in \wp. Appendix A shows
that this simple model provides a surprisingly accurate description of
our numerical results.

\begin{figure}[t]
\centerline{\epsfxsize=3.2truein\epsffile{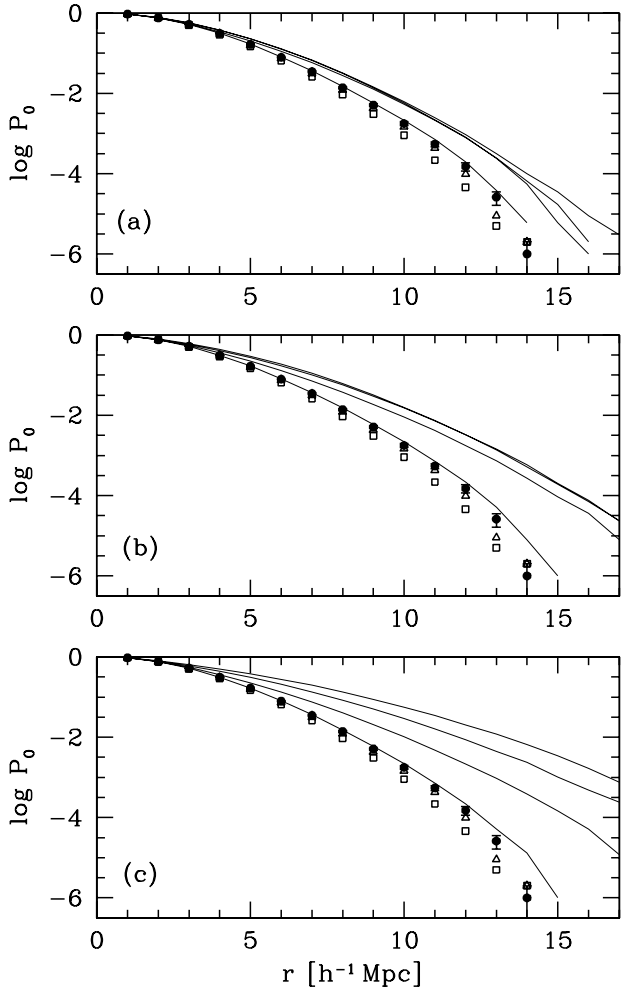}}
\caption{ \label{vpf} Void probability functions for density-dependent
  HOD models of the $M_r<-19$ sample. In all three panels, lines are
  density-dependent HOD models while points show the three models,
  from Figure \ref{vstats}a.  Solid points with error bars represent
  the $(\s8,\slogm)=(0.9,0.1)$ model, open triangles are the
  $(0.9,0.6)$ model, and open squares are the $(0.7,0.1)$ model. Panel
  (a): Models with $\fmin=2$. The four lines, from bottom to top, show
  results for $\dc=-0.8$, $-0.6$, $-0.4$, and $-0.2$. Panel (b):
  Models with $\fmin=4$ for the same values of $\dc$. Panel (c):
  Models with $\fmin=\infty$ for the same values of $\dc$.}
\end{figure}

Figure \ref{vpf} shows the VPFs for $\s8=0.9$ models with the three
values of $\fmin$. In each panel, the filled circles with error bars
show the results for $\slogm=0.1$ from Figure \ref{vstats}. The lines,
from bottom to top, plot $P_0(r)$ for $\dc = -0.8$, $-0.6$, $-0.4$,
and $-0.2$.  (from lowest to highest).  Panel (a) shows the results
for $\fmin=2$. For $\dc=-0.8$ (the lowest line), the effect on $P_0$
is negligible. This is to be expected from Figure \ref{f_delta}
because nearly all halos above $\mmin$ are in regions above
$\delta=-0.8$ anyway. For $\dc\ge-0.6$, the change in $P_0$ is
substantial; at $P_0=10^{-4}$, the change in radius is nearly 2 \hmpc,
while at 12 \hmpc, the change in $P_0$ is $\sim 1$ dex. The VPFs
exhibit little further change above $\dc=-0.6$ because halos above
$2\mmin$ are still common at these densities.  The effect of
environmental dependence on $P_0$ thus ``saturates'' at a density that
we denote $\ds$. In the context of Figure \ref{f_delta}, if $\mmin <
\mstar(\dc) < \fmin\mmin$, then the density dependence eliminates most
halos that would have been able to host galaxies because $\fmin\mmin$
is on the exponential cutoff of the mass function for regions with
$\delta<\dc$. This significantly increases the probability that a
low-density region that was otherwise occupied is now counted as a
void. If both $\mmin$ and $\fmin\mmin$ are below $\mstar$, then the
density dependence moves the minimum mass scale along the power-law
portion of the halo mass function, but not into the exponential
cutoff. The number of halos massive available to host galaxies is
reduced, but the fractional change in galaxy-hosting halos is much
smaller. Thus the likelihood of increasing the sizes of voids is
low. In order to further expand the voids, $\fmin$ must be increased.

Figure \ref{vstats}b plots the results for $\fmin=4$. For $\dc=-0.8$,
the VPF is similar to that of $\fmin=2$, and the VPF again changes
dramatically for $\dc=-0.6$. Because of the higher $\fmin$,
``saturation'' sets in at a higher threshold density $\dc\approx
-=0.4$. At this $\dc$, the VPF is greatly amplified relative to the
density-independent models; the change in the radius at which
$P_0=10^{-4}$ is $\sim 3.5$ \hmpc, and the increase in $P_0$ at 12
\hmpc\ is nearly 1.5 dex. Increasing $\fmin$ further effectively
removes the saturation effect. Panel (c) shows the results for
$\fmin=\infty$. This model evacuates regions below $\dc$ completely,
regardless of halo mass. At a fixed radius the probability of finding
a void increases monotonically with $\dc$.

\begin{figure}[t]
\centerline{\epsfxsize=3.2truein\epsffile{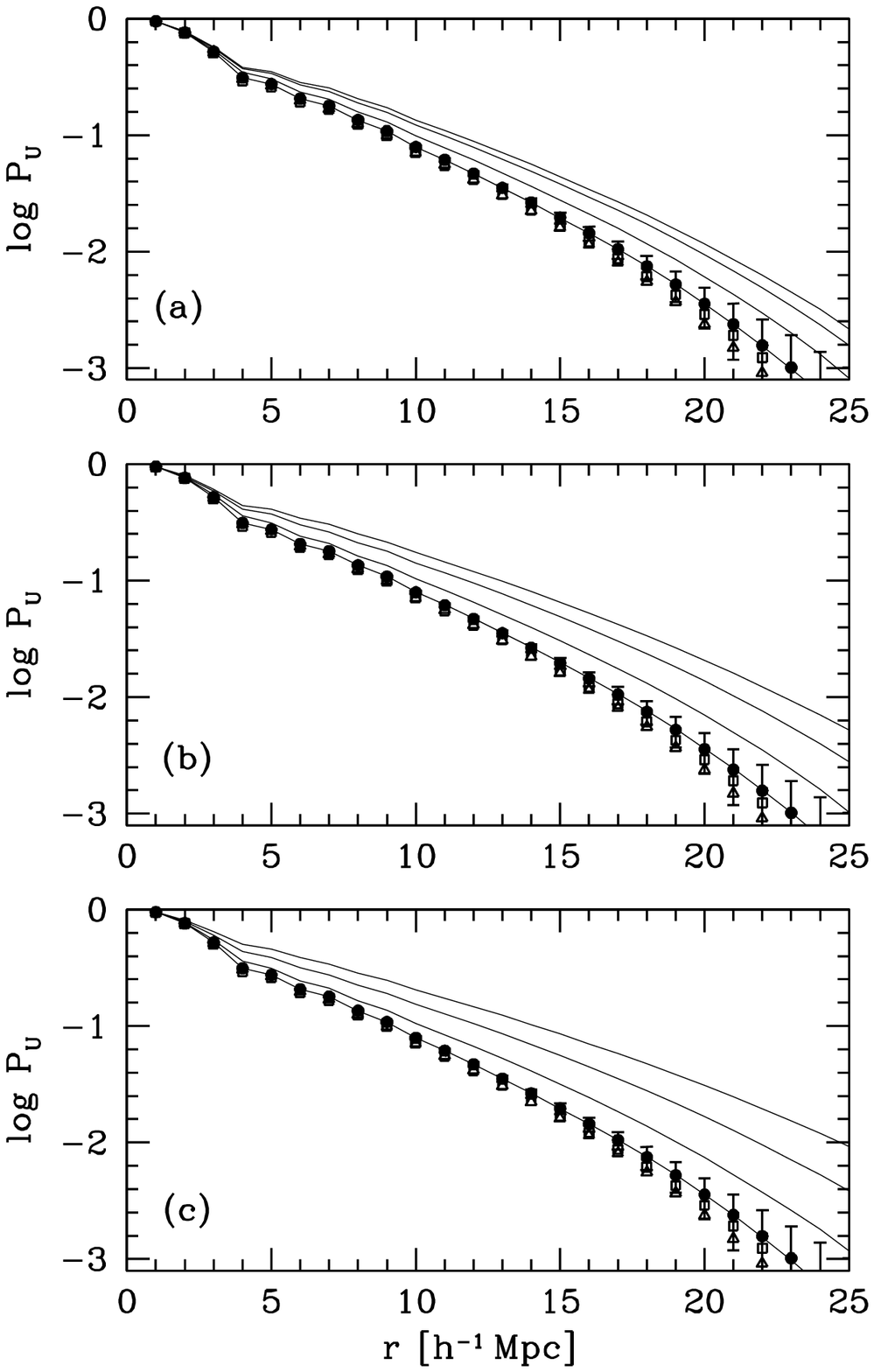}}
\caption{ \label{upf} Underdensity probability functions for
  density-dependent HOD models of the $M_r<-19$ sample. In all three
  panels, lines are density-dependent HOD models while points show the
  three models from Figure \ref{vstats}a.  Solid points with error
  bars represent the $(\s8,\slogm)=(0.9,0.1)$ model, open triangles
  are the $(0.9,0.6)$ model, and open squares are the $(0.7,0.1)$
  model. Panel (a): Models with $\fmin=2$. The four lines, from bottom
  to top, show results for $\dc=-0.8$, $-0.6$, $-0.4$, and
  $-0.2$. Panel (b): Models with $\fmin=4$ for the same values of
  $\dc$. Panel (c): Models with $\fmin=\infty$ for the same values of
  $\dc$.}
\end{figure}

Figure \ref{upf} presents the UPF results for the same models. The
lines and points in each panel correspond to the same models at Figure
\ref{vpf}. Density dependence in the HOD has a different effect on the
UPF than on the VPF. An increase in $\dc$ at a given $\fmin$ always
reduces the number of galaxies in a region of low density, even though
that region may not be a completely empty void. Therefore the
saturation effect seen in the VPF is not as striking for the UPF. In
panel (a), increasing $\dc$ always results in a higher UPF, but the
difference between the models gets smaller at higher $\dc$. As with
the VPF, the increase in $P_U$ is larger for higher $\fmin$. At
$(\fmin,\dc)=(4,-0.4)$, the change in $r$ with respect to the
density-independent models is 5 \hmpc\ for $P_U=10^{-2}$. At $r=20$
\hmpc, the increase in $P_U$ is 0.7 dex.


\subsection{Results for $M_r<-21$}

\begin{figure}
\centerline{\epsfxsize=3.2truein\epsffile{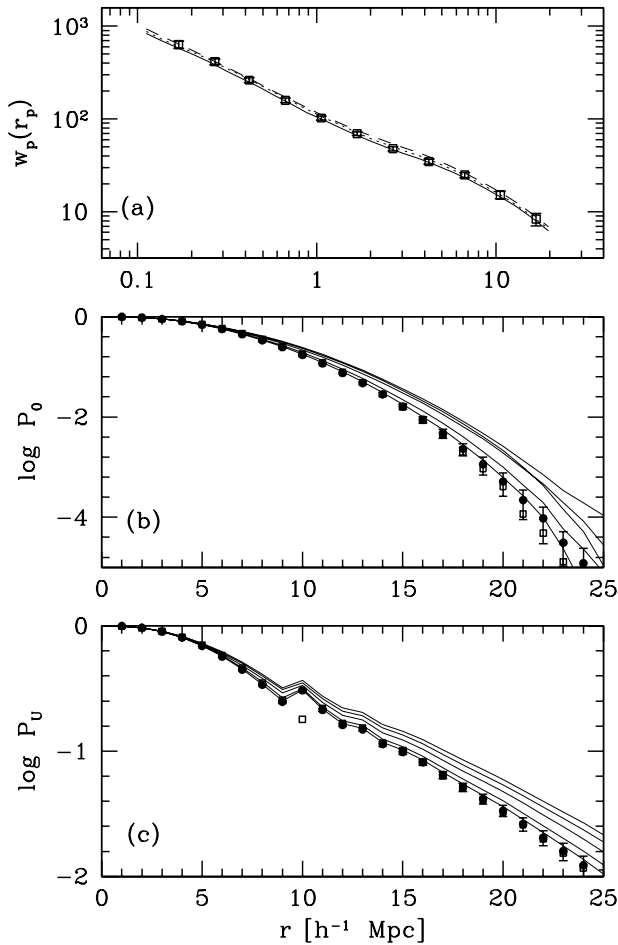}}
\caption{ \label{denstats_210} Clustering statistics for
  density-dependent HOD models of the $M_r<-21$ sample. All models
  have $\fmin=2$. Panel (a): Projected correlation function. Points
  with errors are the SDSS data from Figure \ref{wp_fits}. Line types
  are results for $\dc=-0.2$ ({\it solid}), $\dc=0$ ({\it dotted}),
  and $\dc=+0.2$ ({\it dashed}). Panel (b): Void probability
  functions. Solid points with error are results from Figure
  \ref{vstats}b for $(\s8,\slogm)=(0.9,0.1)$. Open squares plot the
  results for $(0.7,0.1)$. The five lines show results for $\dc=-0.4$
  to $+0.4$ in steps of $0.2$. Panel (c): Underdensity probability
  functions for the same models shown in panel (b). The open squares
  are plotted but cannot be distinguished from the other points. As
  with the $-19$ samples, increasing $\dc$ corresponds to increasing
  VPF and UPF.}
\end{figure}

Our density-dependent HOD models for the brighter, $M_r<-21$ galaxies
employ the same values of $\fmin$, but larger values of $\dc$ are
required to produce a noticeable change in the galaxy
distribution. For conciseness, we present results for $\fmin=2$ only,
but increasing $\fmin$ has the same qualitative effect as shown for
the $M_r<-19$ sample.  Figure \ref{denstats_210}a plots \wp\ for
$\dc=-0.2$, $0$, and $+0.2$. Recall that the density-independent HOD
fits from Figure \ref{wp_fits} were statistically acceptable, but that
the large-scale amplitude of the models was slightly below the
observed level. Boosting \wp\ by adding density-dependence to the HOD
decreases $\chi^2$ for $\dc\sim0$, but the fit becomes poor for
$\dc\gtrsim+0.3$. Figure \ref{denstats_210}b plots $P_0(r)$ for models
with $\dc=-0.4$ to $+0.4$ in steps of 0.2. At $\dc=-0.4$ there is no
change in the sizes of voids. The effect of environmental dependence
does not become significant until $\dc\ge-0.2$. As with the $M_r<-19$
models, the effect on the VPF reaches a saturation limit at high
$\dc$. For $\fmin=2$ the saturation density is $\ds=0$. Figure
\ref{denstats_210}c plots $P_U(r)$ for the same values of $\dc$. As
with the faint galaxy sample, as $P_U(r)$ increases monotonically as
$\dc$ increases. The effect on the UPF also becomes detectable at
$\dc\ge-0.2$.

\begin{figure}[t]
\centerline{\epsfxsize=3.2truein\epsffile{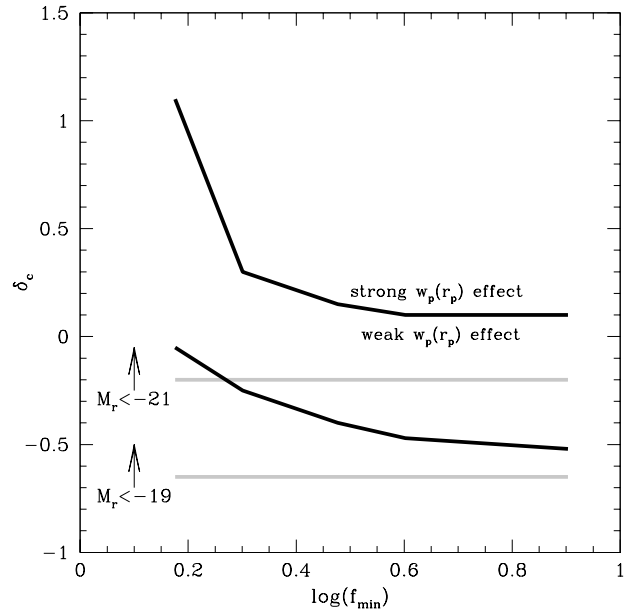}}
\caption{ \label{weak_strong} Regions in the $\fmin-\dc$ parameter
  space where environmental dependence of the HOD has a readily
  measurable impact on void statistics or on the projected correlation
  function, for galaxies with $M_r<-19$ (lower lines) or $M_r<-21$
  (upper lines). The gray lines denote the space above which the
  effect on the void statistics is measurable. The black lines denote
  the space above which the effect on \wp\ is strong ($\Delta
  \chi^2>4$). Between the gray and black lines are the regions of
  parameter space in which environmental dependence is detectable with
  void statistics but does not significantly affect the correlation
  function.}
\end{figure}

Figure \ref{weak_strong} roughly delineates the areas in parameter
space that produce weak and strong effects on galaxy clustering. The
gray lines show the limit where density dependence produces measurable
changes in the void statistics. The lower gray line is for $M_r<-19$
and the upper gray line is for $M_r<-21$. Above the gray lines, there
is a region of parameter space in which the effect on \wp\ is weak,
defined as $\Delta\chi^2<4$ with respect to the best-fit
density-independent model. The black lines denote the transition
between strong and weak influence on \wp. For low $\fmin$, the region
of weak dependence spans a larger range in $\dc$. As $\fmin$
increases, the lines flatten out at $\dc=-0.5$ and $+0.1$ for the
$-19$ and $-21$ samples respectively. Below these densities there are
relatively few halos above $\mmin$; there are enough to alter the
distribution of voids but not to change the overall bias of the galaxy
sample.


\section{Summary and Discussion}

We have examined how non-linear galaxy bias, described in the HOD
framework, affects the distribution of void sizes once the galaxy
number density $\ngavg$ and projected correlation function \wp\
are imposed as observational constraints.  After choosing HOD 
parameters that reproduce these observables as measured for SDSS
galaxy samples with $M_r<-19$ and $M_r<-21$, we compute VPFs
and UPFs by populating the halos of a large, high-resolution
N-body simulation of a $\Lambda$CDM cosmological model.
We consider ``standard'' models in which the galaxy HOD is independent
of environment and models that incorporate a simple form of
environmental variation.

Our parameterization of the HOD incorporates a smoothly truncated
step function for the mean occupation $\ncen$ of central galaxies
and a smoothly truncated power-law for the mean occupation of $\nsat$
of satellites.  Void statistics are sensitive to $\ncen$, since
halos massive enough to host multiple galaxies generally reside
in overdense regions.  If the width of the cutoff in $\ncen$ is
specified, then the cutoff mass scale $\mmin$ is tightly 
constrained by the combination of \wp\ and $\ngavg$, and variation
of $\mmin$ in the allowed range produces almost no change in 
the VPF or UPF.  Broadening the cutoff in $\ncen$ allows some
galaxies to occupy lower mass, less biased halos, slightly
reducing void sizes, but the effects are smaller than the statistical
errors expected from the SDSS.  

More generally, fitting the projected correlation function tightly
constrains the fraction $\fsat$ of galaxies that are satellites, 
with $\fsat \approx 0.23$ and $\fsat \approx 0.15$ for
$M_r<-19$ and $M_r<-21$, respectively, assuming $\sigma_8=0.9$.
Once $\fsat$ is pinned down by \wp, the $\ngavg$ constraint determines the
space density of single-occupancy halos, and because these are
already in the mass regime where spatial bias depends weakly
on halo mass, the void statistics are essentially fixed.
When $\sigma_8$ is lowered to 0.7, the HOD changes required
to increase bias of the correlation function almost exactly 
compensate the reduced void sizes in the matter distribution,
leading to nearly identical galaxy void statistics.
Thus, the assumption of an environment-independent HOD that
reproduces the observed \wp\ and $\ngavg$ leads to a robust
prediction of the VPF and UPF.


Void statistics are, however, sensitive to differences in the 
minimum halo mass scale between low and high density regions.
We have introduced parameterized models that incorporate such
environment-dependent HODs by changing the cutoff mass $\mmin$
by a factor $\fmin$ in regions where the large scale density
contrast $\delta$ (defined with a 5\hmpc\ tophat) is below
a critical value $\dc$.  For large values of $\dc$
(above $\sim -0.5$ for $M_r < -19$ and $\sim 0.1$ for $M_r<-21$,
depending on $\fmin$), density dependence of the HOD changes
the correlation function in a way that cannot be masked by
varying other HOD parameters.  However, there is a substantial
range of $(\fmin,\dc)$ values for which density dependence
of the HOD has a negligible impact on \wp\ but a readily measurable
impact on the VPF and UPF.  Furthermore, the two void statistics 
provide somewhat complementary information, allowing one to 
estimate separately the critical density and the magnitude of the 
shift in $\mmin$.  

Because they are sensitive to environmental variations of the HOD,
void statistics can remove one source of uncertainty in deriving
cosmological constraints from observed galaxy
clustering.  For a specific example, recall that the $\sigma_8=0.7$
model underpredicts \wp\ for $M_r<-19$ galaxies if we assume
that the HOD is independent of environment (Fig.~\ref{wp_fits}a).
Raising $\mmin$ in low density regions could boost the 
galaxy bias factor and thereby improve the agreement between this model
and the SDSS \wp\ measurements.  However, any change large enough
to alter the correlation function at this level should be easily
detected or ruled out by void statistics.
There may be environmental variations in high mass halos that
could affect \wp\ without altering void statistics,
but voids are the ``canary in the coal mine'' for any
changes in the large scale galaxy bias factor driven by
suppression (or enhancement) of galaxy formation in low density regions.

In hierarchical scaling models, the void probability function can be
written as an infinite sum of all $n$-point correlation functions,
which can in turn be expressed as products of the two-point
correlation function (\citealt{white79, fry86}). \cite{conroy05}
argue that the VPF provides little new information about the HOD
once \wp\ has been accurately measured because of this hierarchical
scaling.  While we agree that
the VPF is essentially determined by matching \wp\ for 
environment-independent HOD models, we do not think
that hierarchical scaling alone explains this result.
First, one could create two HOD models with virtually identical
void statistics but very different correlation functions 
simply by moving satellite galaxies to more massive halos
while keeping central galaxy occupations fixed.  Conversely,
our $(\fmin,\dc)$ models include cases with very 
similar correlation functions but substantially different VPFs.
Hierarchical scaling arguments correctly predict the sign of the 
VPF changes in Figures~\ref{vpf} and~\ref{denstats_210} --- models
with higher \wp\ have larger voids --- but they do not
explain the magnitude of these changes.  The relation between
void probabilities and the two-point correlation function depends
not on the hierarchical {\it ansatz} alone but on the specific
amplitudes that relate $n$-point functions to the two-point
function, and these can themselves change with the HOD.

Recent N-body studies show a strong correlation between halo
formation time and large scale environment for low mass halos
\citep{gao05,harker05,wechsler05,zhu06} --- in particular,
the oldest halos avoid low density environments.
At fixed halo mass, an older halo could plausibly host
a more luminous central galaxy because of more efficient 
gas cooling and cannibalism of satellites, or it could host a less 
luminous central galaxy because of greater fading of
stellar populations (see \citealt{zhu06} for example calculations).
Precise measurements of void statistics for the SDSS galaxy samples
modeled here will provide an excellent probe of the link between
halo formation time and $r$-band galaxy luminosity.
A strong correlation between the two would cause significant
departures from the ``standard'' (environment-independent) HOD
predictions presented in Tables~2 and~3, while good agreement
with these predictions would indicate that luminosity is determined 
by halo mass with little dependence on formation epoch.
Extending these arguments and analyses to samples 
of lower luminosity and additional selection criteria such as
color or surface brightness will allow detailed exploration
of the links between a galaxy's present day properties and the
assembly history of its dark matter halo.

\vspace{1cm}

JT would like to thank Charlie Conroy for many useful discussions.  JT
and DW acknowledge support from NSF grant AST-0407125. JT was
supported by a Distinguished University Fellowship at Ohio State
University during the course of this work. Portions of this work were
performed under the auspices of the U.S. Dept. of Energy, and
supported by its contract \#W-7405-ENG-36 to Los Alamos National
Laboratory.  Computational resources were provided by the LANL open
supercomputing initiative.


\vspace{2cm}
\appendix
\section{The Voided Poisson Model}

For a random distribution of points the correlation function is zero at
all $r$. If spherical regions of radius $R_V$ are voided, and the
displaced galaxies randomly distributed through the non-void regions, a
correlation is induced in the points which is non-zero at scales smaller
than $2R_V$. \cite{babul91} presents the resulting correlation function
for a given void radius $R_V$ as

\begin{equation}
\label{e.xi_rv}
1+\xi_V(r,R_V) = \exp \left\{ \frac{f_V}{2} \left[ \left(\frac{r}{2R_V}\right)^3
	- 3\left(\frac{r}{2R_V}\right) + 2\right] \right\}
\end{equation}

\noindent
for $r\le 2R_V$, and $1+\xi(r,R_V)=1$ otherwise. The parameter $f_V$ is
the filling factor of the voids, defined as $f_V=4\pi R_V^3 n_V/3$,
where $n_V$ is the number density of non-overlapping voids.

Originally proposed as a model for the clustering of Lyman-$\alpha$
absorption clouds with ``spheres of avoidance'' carved out by quasars,
this model is also applicable to galaxy clustering in our
density-dependent HOD models. Lowering the efficiency of galaxy
formation in low-density regions expands cosmic voids, as demonstrated
in Figures \ref{vpf} and \ref{denstats_210}. The number density of voids
can be derived from the VPF by

\begin{equation}
\label{e.nv1}
\begin{array}{lll}
n_V(R_V) & = & P_0(R_V)\,V_g^{-1} \\
	 & \approx & P_0(R_V)\,\frac{3\pi^2}{32}\frac{(n^\prime V)^3}{V} \\
\end{array}
\end{equation}

\noindent
where $V$ denotes spherical volume, $V_g$ is the volume within which a
void with radius $R_V$ can be moved without coming into contact with a
galaxy, and

\begin{equation}
\label{e.nv2}
n^\prime(R_V) = \frac{d\ln P_0(R_V)}{4\pi R_V^2 dR_V}
\end{equation}

\noindent
(\citealt{betancort90, patiri04}). The clustering of the voided Poisson
model assumes that the void regions are created from initially filled
regions, while the voids in the density-dependent HOD models are simply
enlarged. For the filling factor required in equation (\ref{e.xi_rv}),
we take the difference between $n_V$ inferred from the density-dependent
model and the density-independent model. This gives us the excess number
density of voids created by the density-dependence. To calculate the
clustering induced by spectrum of voids of varying sizes, we integrate
equation (\ref{e.xi_rv}) over $R_V$;

\begin{equation}
\label{e.int_rv}
1+\xi_V(r) = \frac{\int [1+\xi_V(r,R_V)]\,\frac{dn_V}{dR_V}\,dR_V}
	{\int\,\frac{dn_V}{dR_V}\,dR_V}.
\end{equation}

The clustering boost induced by the voids applies only to the two-halo
term of the correlation; the density of occupied halos in the inter-void
space is increased, but the pairs within a single halo are not
altered. The new two-halo term is

\begin{equation}
\label{e.xi2h_v}
1+\xid^{(V)}(r) = [1+\xi_V(r)][1+\xid(r)],
\end{equation}

\noindent
where $\xid(r)$ is the density-independent two-halo term, which is
calculated analytically.

\begin{figure*}
\centerline{\epsfxsize=5.5truein\epsffile{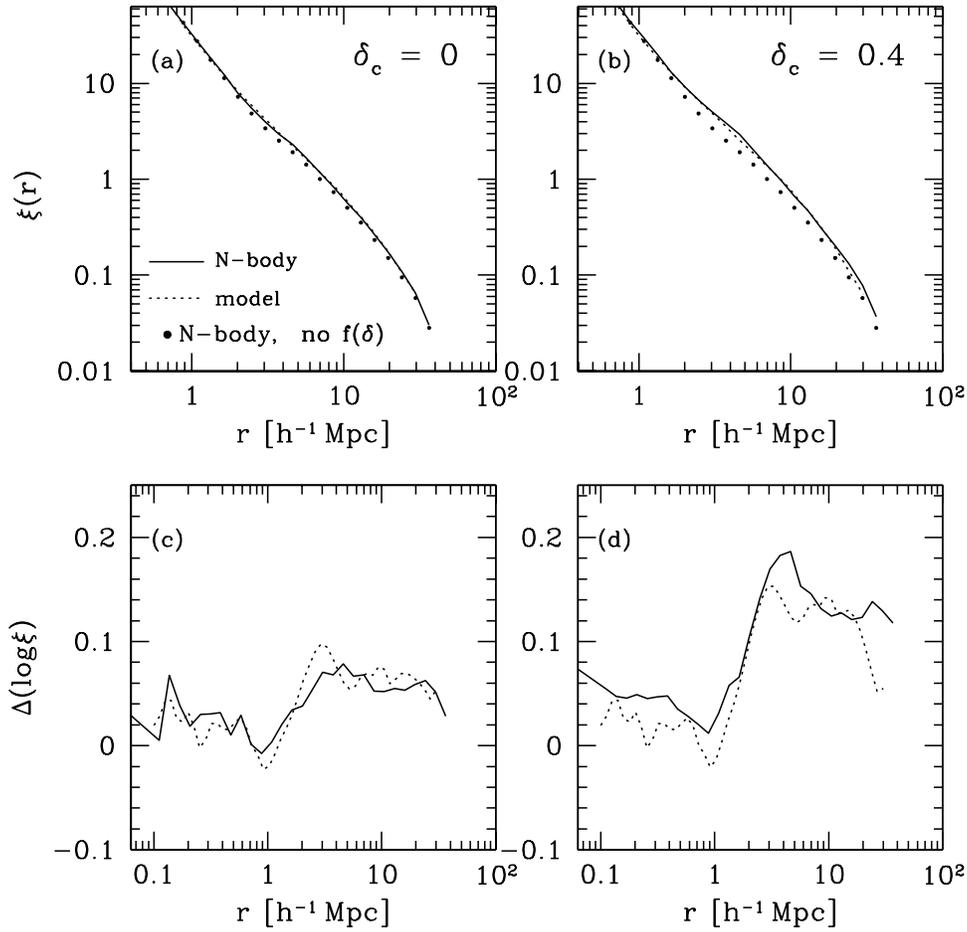}}
\caption{ \label{xi_model} Results of the voided Poisson model of
  Appendix A. All panels show results for models of the $M_r<-21$
  sample with $\fmin=4$. Panel (a): The solid line plots N-body
  results for $\xi(r)$ with $\dc=0$, i.e., $\mmin$ increases in all
  regions below the mean density on the 5 \hmpc\ scale. The dotted
  line shows the analytic calculation. Circles plot the N-body results
  for the model with no density dependence. Panel (b): Same as (a),
  but with $\dc=0.4$. Panel (c): The solid line plots the difference
  in $\log\xi$ between density-dependent N-body results and
  density-independent N-body results from panel (a). The dotted line
  plots the difference between the analytic calculation and the
  density-independent results. Panel (d): Same as (c), but for
  $\dc=0.4$.}
\end{figure*}

Figure \ref{xi_model} shows the results of the model for
density-dependent models of the $M_r<-21$ sample with $\fmin=4$. To
calculate $n_V$, we use the VPFs obtained from the N-body results in
equations (\ref{e.nv1}) and (\ref{e.nv2}). Panel (a) plots $\xi(r)$ for
models with $\dc=0$. The filled circles show the results for the
density-independent model. This value of $\dc$ produces a weak boost in
the large-scale clustering, shown by the solid line. The dotted line,
showing the results of the voided Poisson model, is difficult to see
underneath the N-body results. Panel (b) plots the same results but for
more extreme models with $\dc=0.4$. The effect on $\xi(r)$ is more
pronounced, but it is still well described by the voided Poisson
model. Panels (c) and (d) show $\Delta[\log\xi(r)]$ with respect to the
density-independent model for the density-dependent N-body results and
the model calculations. Panel (d) plots the results for $\dc=0$, which
boosts the large-scale amplitude of $\xi(r)$ by $\sim 0.06$ dex. The
boost abruptly goes away at $r<1.5$ \hmpc, demonstrating that the change
in clustering applies to the two-halo term only. For $\dc=0.4$, shown in
panel (d), the boost is $\sim 0.15$ dex for the large-scale
$\xi(r)$. The model agrees well with the N-body results, but turns over
at $r>20$ \hmpc. This discrepancy is likely a result of truncating the
integral in equation (\ref{e.int_rv}) at 25 \hmpc, which is the limit of
our N-body calculations of the VPF.


The voided Poisson model offers an accurate quantitative explanation of
the effect of our density-dependent HOD models on the correlation
function. Unfortunately, as presented in this Appendix it cannot be used
for full analytic modeling of the effect of density dependence without
the use of N-body simulations, which were used here to obtain $n_V$. The
need for N-body simulations could be circumvented with analytic
approximations of the conditional mass function (\citealt{sheth02,
gottlober03, patiri04}), but it is unclear whether these approximations
have the necessary accuracy for use as a tool for constraining density
dependence with \wp\ measurements. Further investigation may prove
fruitful.




\end{document}